# ENZYME-BASED LOGIC SYSTEMS FOR INFORMATION PROCESSING

**Evgeny Katz[#] and Vladimir Privman[##]**

*Department of Chemistry and Biomolecular*
*Science and Department of Physics,*
*Clarkson University, Potsdam NY 13699, USA*

[#] ekatz@clarkson.edu

[##] privman@clarkson.edu

Click this area for future

updates of the manuscript

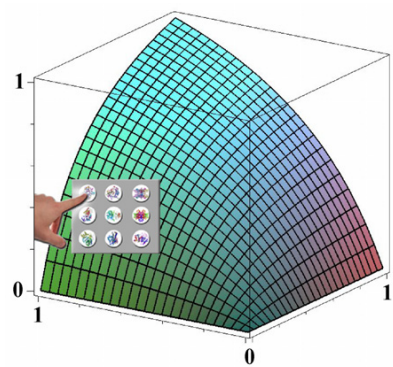

## ABSRACT

We review enzymatic systems which involve biocatalytic reactions utilized for information processing (biocomputing). Extensive ongoing research in biocomputing, mimicking Boolean logic gates has been motivated by potential applications in biotechnology and medicine. Furthermore, novel sensor concepts have been contemplated with multiple inputs processed biochemically before the final output is coupled to transducing "smart-material" electrodes and other systems. These applications have warranted recent emphasis on *networking* of biocomputing gates. First few-gate networks have been experimentally realized, including coupling, for instance, to signal-responsive electrodes for signal readout. In order to achieve scalable, stable network design and functioning, considerations of noise propagation and control have been initiated as a new research direction. Optimization of single enzyme-based gates for avoiding analog noise amplification has been explored, as were certain network-optimization concepts. We review and exemplify these developments, as well as offer an outlook for possible future research foci. The latter include design and uses of non-Boolean network elements, e.g., filters, as well as other developments motivated by potential novel sensor and biotechnology applications.

## 1.    Introduction

This article offers a review of recent advances in the emerging field of biochemical information processing based on logic systems realized by enzyme-catalyzed reactions. Figure 1 offers a schematic of topics covered or commented on in our presentation. We will use "computing" as a short-hand for "information processing." Unconventional computing is a rather broad research field[1] with loosely interrelated topics ranging from quantum computing which aspires to achieve significant speed-up over the conventional electronic computers for some problems, to biomolecular computing with a "soup" of (bio)chemical reactions which are being researched for multi-input, several-step information processing for advanced sensor applications.[2,3] In fact, biochemical reactions are at the core of the mechanism of Life itself, and therefore one could set rather ambitious expectations for how far can (bio)chemical reaction systems be scaled up in complexity, if not speed, for information processing.

The next section (Section 2) offers an overview of the chemical computing research. Our emphasis in the present review is on enzyme-based systems used to realize Boolean gates and small to ultimately moderate and large networks composed of gates and other, non-Boolean elements (to be specified later), with sensor applications in mind. Therefore, in Section 3 we briefly outline some considerations of the information processing paradigms for network fault-tolerance and scalability, and we point out that while biomolecules offer specificity and functionalities designed by Nature, we do not attempt to use them to duplicate cell processes (thus, we avoid the issues of "artificial life" here). Rather, the gate and network designs presently being researched, in many ways parallel the electronic computers with their digital information-processing approach.

Chemical[4] and biomolecular[5] (biochemical) computing involve processing of input signals, as well as the "gate machinery" that is expected and experimentally observed to have a relatively large level of noise, at least several per-cent on the scale of variables normalized to the digital **0** to **1** range of values. Therefore, considerations of avoiding noise amplification by gate and other network element design, and of error correction by using proper network architecture, are quite important even for small networks. We outline aspects of error control in Section 3, cf.

Figure 1. However, we do not address the issue of stochastic sources of noise (see Figure 1) in such reactions, mostly because this subject is not well studied specifically for enzymatic reactions, whereas considerations of the origins of noise generally in chemical reactions would take us into topics in statistical mechanics which are outside the scope of this review. Most of the noise-control related considerations presented (Section 3), apply, in fact, to many other "unconventional computing" systems.

The rest of the review is devoted primarily to the enzyme-based and related information processing ("computing"). In Section 4, we illustrate how Boolean gates are realized in such systems. Section 5 is devoted to analog error suppression by gate design for scalability/robustness/fault-tolerance, whereas some aspects of network optimization are presented in Section 6. In Sections 5 and 6, we also address challenges with semi-quantitative rate-equation modeling of biochemical processes in the context of understanding and optimizing gate functions for enzyme-based computing, which is an important topic on its own (see Figure 1) because of the complexity of typical enzymatic reactions. Section 7 addresses the interfacing of the output signal with stimuli-responsive materials, while Section 8 outlines the coupling of the enzyme logic systems with switchable electrodes and bioelectronic devices (e.g. biofuel cells). Finally, Section 9 offers a summarizing discussion of long-term challenges in this field of research and applications.

## 2. From Chemical to Biomolecular Computing

Chemical computing, as a research sub-area of unconventional computing, aims at using molecular or supra-molecular systems to perform various computing operations mimicking processes typical for electronic computing devices.[4] Chemical reactions observed as changes of bulk material properties or structural re-organizations at the level of single molecules can be described in the information processing language, thus allowing formulation of chemical processes in terms of computing operations rather than traditional for chemistry materials transformations.[6] Many of the chemical systems used for information processing are based on molecules or supra-molecular ensembles existing in different states reversibly switchable from one to another upon application of various external physical or chemical inputs.[7] Chemical transformations in switchable molecular systems used for mimicking computing operations can be based on redox changes,[8] acid-base[9] or chelating reactions,[10] and isomerization processes.[11] Chemical reactions in switchable systems can be induced by external physical signals (e.g., light,[12] magnetic field,[13] or electrochemical potential[14]) and, sometimes, by chemical signals (e.g., pH changes[15] or metal cation additions[16]). Some of the studied switchable systems can respond to two kinds of physical or physical/chemical signals (e.g., potential applied to an electrode and illumination,[17] or pH change and illumination,[18] or ion addition and applied potential[19]). The output signals generated by the chemical switchable systems are usually read by optical methods (absorbance[20] or fluorescence[21] spectroscopy) or by electrochemical means (currents or potentials generated at electrodes or in field-effect transistors[22]).

Switchable chemical systems can be simple solutions of molecules[23] responding to external physical or chemical signals, or systems assembled at interfaces[2,24] (frequently at surfaces of conducting electrodes or Si-chips electronically communicating with the support). Ingenious supra-molecular ensembles operating as molecular machines with translocation of their parts upon external signals were designed and used to operate as chemical switchable elements performing logic operations.[25] Based on molecular switchable systems various Boolean logic operations such as **AND**,[26] **OR**,[27] **XOR**,[25,28] **NOR**,[29] **NAND**,[30] **INHIB**,[31] **XNOR**,[32] etc., were realized. Sophisticated molecular designs have allowed reversible,[33] reconfigurable[34] and resettable[35] logic gates for processing chemical information. Other chemical systems mimicking various components of digital electronic devices were designed, including molecular comparator,[36] digital demultiplexer,[37] encoder-decoder,[38] keypad lock,[39] as well as flip-flop and Write/Read/Erase memory units.[40] Chemical systems can solve computing problems at the level of a single molecule[41] resulting in nano-scaling of the computing units[42] and allowing parallel computations performed by numerous molecules involved in various reactions.[43]

Despite the fact that chemical computing is a rapidly developing area of research, the field is still in very early experimental and theoretical stages, though great future potential has been envisioned.[44] Chemical (molecular) computing is not considered as a competitor for conventional silicon-based computing, but rather a novel approach to selective applications.

Microrobotics and bioimplantable computing systems are among the most likely applications to benefit from advances in molecular computing. Future progress in these areas will depend on the development of novel computing concepts and design of new signal-responsive and information processing materials contributing to molecular information technology.[45]

One of the most important challenges in chemical computing field is scalability of the systems. Impressive results have recently been achieved in this direction.[4] Combination of chemical logic gates in groups or networks resulted in simple computing devices performing basic arithmetic operations[46] such as half-adder/half-subtracter[47] or full-adder/full-subtracter.[48] Integration of several functional units in a molecular structure resulted in multi-signal responses to stimuli of various chemical or physical natures, thus allowing different logic operations or even simple arithmetic functions to be performed within a single polyfunctional molecule.[49] Artificial abiotic systems mimicking elementary properties of neuron networks were developed demonstrating that scaling up the complexity of chemical computing systems can be inspired by biological principles.[50] However, most of the recently developed chemical computing systems[4] operate as single logic gates without possibility to be concatenated in networks. In many systems the physical nature of the input and output signals is different, making the assembly of multi-component logic networks difficult or impossible. Complex multi-component chemical logic systems usually require ingenious supramolecular ensembles to allow compatibility of the information processing sub-units.[4] Supramolecular systems functioning as molecular machines have been designed and used to operate as chemical computing elements, but they performed rather simple functions despite their extreme synthetic complexity.[51]

Many of the problems hardly addressable by synthetic chemical systems can be solved naturally by utilization of biomolecular systems. The emerging research field of biocomputing, based on application of biomolecular systems for processing chemical information, has achieved higher complexity of information processing while using much simpler chemical tools, due to the natural specificity and compatibility of biomolecules.[52] Different biomolecular tools, including proteins/enzymes,[53] DNA,[54] RNA,[55] and whole cells[56] were used to assemble computing systems processing biochemical information. Various Boolean logic operations were mimicked by enzyme systems[57,58] allowing concerted operation of multi-enzyme assemblies performing simple arithmetic functions (half-adder/half-subtracter).[59] Although similar logic operations and arithmetic functions were also realized using non-biological chemical systems,[46] the advantage of the biomolecular systems has been in a relative simplicity of the various assembled logic schemes. The complexity of information processing systems with enzymes could be easer scaled up, resulting in artificial biocomputing networks performing various logic functions (e.g., implication) and mimicking natural biochemical pathways.[60,61] Some of the designed biomolecular computing assemblies (particularly DNA-based biocomputing systems[62]) have been devised for solving complex combinatorial problems, promising to have an advantage over silicon-based electronic computers due to parallel computing performed by numerous biomolecular units. However, the main advantage of biomolecular computing systems over electronic computers is their ability to process biochemical information received in the form of chemical inputs directly from biological systems with the possibility to operate in a biological environment (biomedical/diagnostic applications).[63]

## 3.   Information Processing Paradigms and Control of Noise for Scalability

Figure 2 offers a schematic of the presently known/researched paradigms of information processing that are expected to allow "scalability" for large-network, complex functioning with fault-tolerance, i.e., with the ability to devise the system's architecture to control buildup of noise with non-prohibitive resources. We intentionally use "non-prohibitive" instead of "non-exponential," etc., because we do not want to go into the mathematical definitions of concepts such as fault-tolerance. Furthermore, some of the items in Figure 2 will probably be questioned by purists on the grounds of not being proven scalable to indefinitely large complexity, or simply not yet well understood, such as Life. Indeed, all the information processing paradigms listed are presently actively researched. Specifically, processes in living things are the subject of Systems Biology.[64] Quantum computing has been an explosively growing field of research with presently only very limited systems of few qubits (quantum bits) realized.[65] DNA and certain other biomolecule-based computing is a subfield of biomolecular information processing specifically relying on various forms of "ensemble" parallelism.[62]

The enzyme-based computing approach reviewed here, follows the conventional information processing paradigm of modern electronics: digital[66] approach based on analog gates and other elements operating in a network. The reasons for this choice have been several-fold. Firstly, mimicking natural processes in a cell requires an extremely complicated set of coupled biochemical reactions. Enzymatic computing is presently far from the complexity needed to attempt "artificial life" realizations. Secondly, most applications of enzyme-based systems are expected to be in novel sensing approaches[3] involving processing of several input signals. The ultimate outcome in these systems is of the Yes/No type, corresponding to "Sense/Act" or "Sense/Diagnose/Treat." Thus, the output is by definition digital, and at some point, either in the biochemical stages or during signal transduction to electrodes/electronic computers for the "Act" ("Treat") step, the Yes/No digitization will be imposed.

Thirdly, and most importantly, the use of the digital information processing paradigm for enzyme-based computing offers a well established approach for control of the level of noise buildup in networks of biochemical information processing reactions. Chemical and biochemical systems are much more prone to noise than electronic computer components: Their applications are in environments where the inputs (reactant chemicals' concentrations) and the "gate machinery" (other chemicals' concentrations) are all expected to fluctuate within at least a couple of percent of the range of values between the "digital" $0$ and $1$. Therefore, consideration of control of noise is required already in concatenating as few as 2-3 gates.[3,67–69]

We note that digital information processing is actually carried out by network elements which are analog in nature. This is illustrated in Figure 3. Figure 3(a) shows the simplest possible logic function: an identity "gate" with one input and one output. However, the actual system response in various applications can be, for instance, of the type shown in Figure 3(b-c). These are the simplest possible concave and convex response curves. The details of the response, especially the linear regions, are important in certain sensor applications. In fact, more complicated shapes are possible; cf. Figure 3(d). However, in most catalytic biochemical reactions the convex response curve, Figure 3(c), is found, because the output signal — a product of the reaction — is controlled by and typically proportional to the input-signal chemical concentration for small inputs. However, for large inputs, the output is usually limited for example by the reactivity of the available biocatalyst. Therefore the curve reaches saturation.

While most biochemical gates realized have two inputs, as described in Sections 4-8 below, the above single-input discussion provides a useful background to outline the issue of controlling noise amplification. Indeed, consider Figure 4, which shows a typical actual signal value distribution (peaked at 1) in the situation when the digital result expected is **1** (with a similar distribution for digital **0**, peaked at 0, and, for situations when this value is also the physical zero of a chemical concentration, spread to positive values only). By "analog" noise we denote the mere spread of the values in the range about 1, which for chemical computing is unfortunately rather broad, typically at least a couple of percent. In order not to broaden this spread, we have to pass the signal through "filters" of the type shown in Figure 3(d). In fact, ideally we would like such a sigmoid property (small slopes/gradients at and near the digital points) to be present in all or most of our gates. Filters can also be used as separate elements/steps. There is evidence that both of these solutions for suppressing analog noise buildup are present in Nature.[70]

However, filtering can push values which are away from the correct digital result (in our case, values in the tail of the distribution close to 0, see Figure 4, i.e., far from the expected 1) to the wrong answer (here, to 0). Thus, the process of digitization itself introduces also the "digital" type of noise. However, such errors are not very probable and only become important to actively correct for larger networks. Standard techniques based on redundancy are available[71] for digital error correction. In enzyme-based computing, for the presently realized network sizes and levels of noise, it is the analog error correction that is important and has recently received significant attention,[2,3,67–69,72,73] as reviewed in Sections 5 and 6.

## 4.    Examples of Boolean Gates Involving Enzyme-Catalyzed Biochemical Reactions

Various logic gates were designed with the use of enzyme reactions: **XOR**, **INHIBIT A**, **INHIBIT B**, **AND**, **OR**, **NOR**, **Identity** and **Inverter**.[58,59] In the most studied experimental approaches, the enzyme systems carrying out Boolean logic operations were used in solutions, performing simple biocatalytic reactions. Recently, experiments have been reported with enzymes as a "gate machinery" immobilized on solid supports[57,74] or at interfaces [2,75] allowing their separation from the reacting species. Our examples below are of the former, "chemical soup in solution" type. Note that enzymes have been used not only as the biocomputing gate "machinery" processing chemical signals,[58,59] but also as the input signals activating a "soup" of chemicals.[57]

In the examples below we follow the convention commonly practiced for enzyme-based computing systems, that the logic **0** and **1** are represented, respectively, by the lower (in this section, physical zero) and higher (bio)chemical concentrations. The actual values for "model" experiments are conveniently selected to have reasonable time scales and signal intensities. However, in applications, many (bio)chemical concentrations are actually pre-defined, for example by the physiological concentrations, and the role of **0** and **1** may have to be redefined appropriately.

For illustration, let us consider a two-enzyme system composed of soluble glucose oxidase (GOx) and catalase (Cat) constituting the logic gate "machinery" which was activated with two chemical input signals: $H_2O_2$ and glucose; see Figure 5(A).[58] The product of the biocatalytic oxidation of glucose — gluconic acid — was detected by optical means in the presence of special "developing" reagents, and the resulting optical absorbance change, $|\Delta A|$, was considered as the output signal of the gate, Figure 5(B). The oxidation reaction required the presence of glucose and $O_2$, where the latter was produced *in situ* upon $H_2O_2$ reacting with Cat. Thus, the output signal appeared only in the presence of both inputs (combination **1,1**), while in the absence of any or both input signals (combinations **0,1**; **1,0**; **0,0**, respectively) the system did not produce gluconic acid. In order to define digital values of the output signal, a threshold value separating small background optical changes defined as **0** output and large absorbance changes defined as **1** output was used, Figure 5(C). The features of this biocatalytic system resemble the Boolean **AND** logic gate.

As another example, let us illustrate the Boolean **XOR** logic operation performed by enzymes. Here glucose dehydrogenase (GDH) and horseradish peroxidase (HRP) produce NADH and NAD$^+$ in the presence of glucose and H$_2$O$_2$, respectively, Figure 6(A).[58] The system has two different reaction pathways: reductive and oxidative, biocatalyzed by GDH and HRP, respectively. When both of them are balanced (input signals combinations **0,0** or **1,1**) the ratio of the reduced and oxidized cofactor is not changing. However, unbalancing the reactions results in the increase or decrease of the NADH concentration, thus affecting its optical absorbance, see Figure 6(B). When the output signal was defined as the absolute value of the absorbance changes, the system demonstrated the features of the Boolean **XOR** logic gate, Figure 6(C).

The above examples are based on simple biocatalytic reactions. Additional illustrations of enzyme-based gates will be given in Sections 7 and 8. Upon scaling up the information processing reactions by concerted operation of many enzymes the complexity can be substantially increased. Coordinated operation of several enzyme-based logic gates was demonstrated,[60] revealing new properties of information processing by enzyme reactions. For example, a biocomputing security system was designed mimicking a keypad lock device, where the final output signal "YES" was achieved only when all chemical input signals were applied in the correct order (**Implication** logic function).[61] By combining the **AND** and **XOR** or the **XOR** and **INHIBIT A** gates, the half-adder and half-subtractor were constructed, respectively, opening the way to elementary computing with enzymes.[59] For example, a sequence of biochemical reactions biocatalyzed by GDH, HRP, GOx and Cat, activated by the additions of glucose and H$_2$O$_2$ can produce a simple arithmetic function operating as a half-adder.[59] Two chemical input signals represented the first and second digits to be summed, while the two output signals were the carry and sum digits of the result, Figure 7. Similarly, a half-subtractor has been realized.[59]

## 5.    Modeling of Enzymatic Reactions and Gate Design for Control of Analog Noise

In this section, we consider how and to what extent can a single two-input, one-output enzymatic gate be analyzed and optimized for avoiding the amplification and preferably actually suppressing the spread of the distribution of the signals about their digital values. For definiteness, we consider **AND** gates; and we will focus on a specific example. Let us first address a generalization of the discussion in connection with Figure 3, to two-input gates. This is presented in Figure 8, which shows a typical biocatalytic gate function with the initially linear increase in response to both inputs, and ultimate saturation, as well as the desired, but difficult to realize with a single biocatalytic reaction, sigmoid response in both variables. Two recently realized gate functions[2,73] which allow avoiding noise amplification, as detailed later, are also shown.

We begin our discussion by considering as an example the **AND** gate realized by two specific chemicals as inputs into a particular enzyme-catalyzed reaction, Figure 9. The substrate, hydrogen peroxide (H$_2$O$_2$), of time-dependent concentration $I_1(t) = [\text{H}_2\text{O}_2](t)$, constitutes the first input at time 0, $I_1(0) = [\text{H}_2\text{O}_2](t = 0)$. It reacts with the biocatalyst: here enzyme horseradish peroxidase (HRP), of concentration to be denoted $E(t) = [\text{HRP}](t)$. The produced complex of concentration $C(t)$, then reacts with the co-substrate 2,2'-azino-*bis*(3-ethylbenzthiazoline-6-sulphonic acid) (ABTS), the resulting oxidized form of which is optically detected. Note that actually two molecules of ABTS are consumed for each H$_2$O$_2$, but due to the rate differences, one of the steps is rate-determining, as commented on later. Thus, the co-substrate of concentration $I_2(t) = [\text{ABTS}](t)$, represents the second input signal at time 0, $I_2(0) = [\text{ABTS}](0)$, whereas the concentration of it's oxidized form, which is the reaction product, to be denoted $P(t) = [\text{ABTS}_{\text{ox}}](t)$, detected at a certain time $t^{\text{gate}}$, is the output signal $P(t = t^{\text{gate}})$.

Ideally, the gate should be operated with all the input/output chemical concentrations at logic-**0** or **1** values. However, due to noise, signal values not precisely at 0 or 1 are also realized. We define the dimensionless "logic" variables

$$x = \frac{I_1(0)}{I_1^{\text{gate}}}, \quad y = \frac{I_2(0)}{I_2^{\text{gate}}}, \quad z = \frac{P(t^{\text{gate}})}{P^{\text{gate}}}, \qquad (1)$$

where $I_1^{\text{gate}}$, $I_2^{\text{gate}}$, $P^{\text{gate}}$ are identified as the logic-**1** reference values. Note that the inputs $I_1^{\text{gate}}$, $I_2^{\text{gate}}$ might be set by the application: The environment of the gate sets those input values $I_1(0) = [\text{H}_2\text{O}_2](t=0)$ and $I_2(0) = [\text{ABTS}](0)$ which constitute logic-**1** (we assume that logic-**0** is at zero). However, the output $P^{\text{gate}}$ is set by the gate-function itself (up to a possible noise in it, as mentioned later, implying that $P^{\text{gate}}$ is, for instance, defined as the mean value).

The output of the gate, $P(t^{\text{gate}})$, is a function of the inputs, $I_{1,2}(0)$, but also of many other parameters, such as the initial enzyme concentration, $E(0)$, the gate time, $t^{\text{gate}}$, and the properties of the reaction kinetics: rate constants, e.g., $r$, $R$, etc., which in turn depend on the chemical and physical conditions of the gate's environment. For the logic-implementation analysis, we seek a parameterization of the gate-response function when it is expressed in terms of the normalized variables in Eq. (1),

$$z = F(x,y) = F(x,y; E(0), t^{\text{gate}}, r, R, \ldots ; \ldots). \quad (2)$$

This function should be considered for general $x, y, z$ ranging from 0 to 1 (and to values somewhat larger than 1). The second expression in Eq. (2) emphasizes that the gate response function also depends on the potentially controllable parameters mentioned earlier, $E(0)$, $t^{\text{gate}}$, $r$, $R$, … that can in principle be adjusted to improve the gate performance, as well as on those parameters (marked by the second "…") which are externally fixed, such as additional chemicals in the environment. As mentioned, the function $F(x,y)$ can also have a stochastic noise component in it, but we will ignore this issue for the moment.

There are two aspects of the gate-response function that should be addressed for optimization. One involves the implications of its shape for noise buildup in the system. However, another, no less important matter is the development of modeling approaches to enzymatic reactions with would yield few-adjustable-parameter representations for the function $F(x,y)$. We will discuss the latter aspect presently, with our specific example as an illustration.

Quality of experimental data for mapping out the response surface is illustrated in Figure 10. Typically, the available information (the number of data points) is limited, even for somewhat more detailed mappings as illustrated below, because of the issues of enzymes' stability for the duration of the experiment and reproducibility of their properties (variations in activity). Therefore, a rule of thumb has been to seek few-parameter model representations[2,3,67–69,73] for the gate-response surface. For example, for the present reaction we will use the following simplified reaction scheme,

$$E + I_1 \xrightarrow{r} C \,,$$ (3)

$$C + I_2 \xrightarrow{R} P + E \,.$$ (4)

This two-parameter model involves two rate constants. The model obviously entails several approximations. Specifically, the actual pathways for the present biocatalytic system are rather complicated[76] and not fully understood. They involve several intermediate compounds. We already mentioned earlier that even our definition of the "complex" $C$, is not precise because the intake of ABTS is two-molecule (two-step) albeit with rates different enough so that one of them can be approximately regarded as rate-determining. The latter assumption has allowed us to write a single Eq. (4). Similarly, Eq. (3) involves the assumption of no back reaction (which, if included, would yield the standard Michaelis-Menten model). This seems appropriate[73] in the regime of the present experiment, due to the fact that the rate $r$ is large.[77]

Whatever the specific arguments are used for zeroing in on the key reaction pathways and rate parameters for each particular gate-realizing system, the available experimental data are simply not detailed and consistent enough for a multi-parameter fit of the full complexity of the enzymatic reactions involved. Thus, the approach taken has been that, for the logic-function analysis, we need a semi-quantitative representation of the features of the shape of the response surface, in most cases accomplished by a two-, at most three-parameter model. The actual kinetic rate equations[73] corresponding to Eq. (3–4), are derived in the standard way and are not detailed here. For our HRP-catalyzed reaction, the data fit according to Eq. (3–4) is shown in Figure 10.

When rescaled to the logic-variable range [0,1], the experimental and fitted surfaces in Figure 10 are obviously of the typical biocatalytic type alluded to in Figure 8(A). Analog noise is in general identified as a certain spread in the inputs, $x$, $y$, about 0 or 1. The output, $z$, is then also not precisely at 0 or 1. If we assume that the widths of the input distributions, $\delta x$ and $\delta y$, are small and comparable to each other, $\delta x \approx \delta x \ll 1$, then the resulting deviation, $\delta z$, for smoothly varying gate functions such as the one in Figure 10, can be estimated as

$$\delta z \approx |\vec{\nabla} F| \, \delta x + \delta F \, . \qquad\qquad (5)$$

Here $|\vec{\nabla} F|$ is the magnitude of the gradient vector of the gate function at the appropriate logic point: **00**, **01**, **10**, or **11**. In addition, if the gate function itself is somewhat noisy, with spread $\delta F(x,y)$ about the average $F(x,y)$, then this contribution should be added at each logic point, as shown in Eq. (5). If we ignore the latter effect, then depending on the value of the largest of the four logic-point gradients, $|\vec{\nabla} F|_{\max}$, the gate can amplify the noise level ($\delta z > \delta x, \delta y$), suppress it ($\delta z < \delta x, \delta y$), or keep it approximately constant ($\delta z \approx \delta x, \delta y$).

The best-case scenario (not yet experimentally realized for enzyme-based gates) is, of course, a noiseless (negligible $\delta F$) function $F(x,y)$ with a "sigmoid" shape in both $x$ and $y$ variables, Figure 8(B), will all the gradients smaller than 1 (thus suppressing analog noise). Typically, however, enzyme-based gates, both those with enzymes as the "machinery" and those with enzymes as inputs, for conveniently selected experimental parameters show response of the convex type, Figure 8(A), with the largest gradient well over 1. In fact, values for $|\vec{\nabla} F|_{\max}$ as large as ~5 (means, 500% noise amplification factor) are common.

The difficulty with gate optimization for the typical convex shape, Figure 8(A), can be clarified as follows. Dependence on some parameters that control the intensity of the reaction output, specifically, the initial enzyme concentration, $E(0)$, and the reaction time, $t^{\text{gate}}$, is approximately linear over a broad range of reaction conditions and therefore largely cancels out when rescaling the output to the "logic" range [0,1]. Specifically, biocatalytic reactions are frequently utilized in the regime of approximately steady-state[78] output, i.e., linear dependence on $t^{\text{gate}}$. Thus, in order to adjust the shape of the response function, variation of these parameters over several orders of magnitude is needed, which is not practical. Variation of other parameters, such as reaction rates, even when possible, encounters a different type of difficulty. Indeed, reaction rates generally depend on the physical and chemical conditions of the system. However, this dependence for enzymatic reactions is not well understood theoretically or studied experimentally. For example, the fit of the data in Figure 10, collected for $t^{\text{gate}} = 60\,\text{sec}$, with the initial $E(0) = 0.5\,\text{nM}$, yielded estimates for the rate constants: $r = 18\,\mu\text{M}^{-1}\text{s}^{-1}$ and $R = 5\,\mu\text{M}^{-1}\text{s}^{-1}$. With this parameterization, the model has allowed a scan over a range of possible $E(0)$ and $t^{\text{gate}}$ values, which, however, suggested[69,73] that getting the noise amplification factor below about 300% was impractical.

Even when parameter adjustment is possible, empirical evidence has been accumulated[2,3,67–69,73] suggesting that typical convex response-surface shapes can be at best optimized to have the largest gradient somewhat below 1.2, i.e., 20% added noise per each gate function. This implies[67] that without additional network features or elements, such as dedicated filters, and even if we due to the gate function itself, $\delta F$ in Eq. (5), then only about 10 gates can be connected in a network before useful signal will be completely lost to noise. Interestingly, recent experiments have suggested a similar picture for networking of neurons.[79]

Returning to our example involving the enzyme HRP, we note that it is known to take on a variety of co-substrates[77] with a large variation in the rate $R$. Our earlier system corresponded to comparable (in order of magnitude) rates $R$ and $r$. We will now explore a system for which $R << r$: Instead of changing the physical or chemical conditions, which leads to rate variations which are not well studied, we simply select an appropriate co-substrate known to have a much smaller ratio $R/r$, i.e., ferrocyanide. The product of the gate, serving as the output signal (optically detected) is then ferricyanide.[73] The experimental data are shown in Figure 11. The theoretical fit yielded the reaction rate estimates[73] $r = 17\,\mu\mathrm{M}^{-1}\mathrm{s}^{-1}$ and $R = 32 \cdot 10^{-3}\,\mu\mathrm{M}^{-1}\mathrm{s}^{-1}$, consistent with earlier studies.[77] Note that this particular experiment was done with $E(0) = [\mathrm{HRP}](t = 0) = 0.05\,\mu\mathrm{M}$, $t^{\mathrm{gate}} = 60\,\mathrm{sec}$. Interestingly, while the data are well fitted by the rate equations corresponding to same reaction scheme, Eq. (3–4) with $I_2(t) = [(\mathrm{Fe(CN)}_6)^{4-}](t)$, the gate-function response surface is actually not smooth in this case. Rather, is has a rounded ridge reminiscent of, but less symmetrically positioned than that in Figure 8(C).

For gate-function response surfaces of the type shown in Figures 8(C) and 11, as well more generally for relatively wide noise distributions, analysis slightly more complicated than just calculating the maximal gradient at the four logic points is required. To estimate noise amplification,[67,73] we study the width, $\sigma_{\mathrm{out}} = \sigma_z$, of the output signal distribution as a function of the width of the input noise distributions assumed equal for simplicity, $\sigma_{\mathrm{in}} = \sigma_x = \sigma_y$. The specific numbers below are for $\sigma_{\mathrm{in}} = 0.1$. Furthermore, we assume uncorrelated, Gaussian input noise distributions, $G_{\mathbf{0}\,\mathrm{or}\,\mathbf{1}}(x)$, with half-Gaussian for $x$ at logic-$\mathbf{0}$ and full Gaussian at logic-$\mathbf{1}$, and similarly for $y$. The output distribution width is then estimated from $\sigma_{\mathrm{out}}^2 = \langle z^2 \rangle - \langle z \rangle^2$ for logic-$\mathbf{1}$ (and $\sigma_{\mathrm{out}}^2 = \langle z^2 \rangle$ for logic-$\mathbf{0}$), with the moments such as $\langle z^2 \rangle$ of the gate response function $z = F(x, y)$ computed with respect to the product input distribution $G_{\mathbf{0}\,\mathrm{or}\,\mathbf{1}}(x)G_{\mathbf{0}\,\mathrm{or}\,\mathbf{1}}(y)$.

The computation yields spread of the output signal near the respective logic outputs $\mathbf{0}$ or $\mathbf{1}$ for the four logic input combinations $\mathbf{00}$, $\mathbf{01}$, $\mathbf{10}$, and $\mathbf{11}$. In general, one would want to have the maximum of these spreads, $\sigma_{\mathrm{out}}^{\mathrm{max}}$, as small as possible. In fact, for network scalability the actual value of the noise spread is not as important as the degree of noise *amplification* at each gate, measured by $\sigma_{\mathrm{out}}^{\mathrm{max}} / \sigma_{\mathrm{in}}$. (This gate quality measure may somewhat depend on the choice of $\sigma_{\mathrm{in}}$; see, e.g., Ref. 2.) As described earlier, the fact that our "gate machinery" enzyme HRP is not expected to have a double-sigmoid gate-function shape, Figure 8(B), limits the optimized-gate $\sigma_{\mathrm{out}}^{\mathrm{max}} / \sigma_{\mathrm{in}}$ to values somewhat larger than or equal to 1.

The "ridged" area of the response surfaces such as those in Figures 8(C) and 11, has a larger-than-1 gradient but limited angular spread. When a two-variable distribution (the product of the two Gaussians) is mapped onto a single-variable one, the "weight" of this angular region is small. Indeed, detailed analysis[73] of the data in Figure 11, suggest that in fact the quality measure $\sigma_{out}^{max} / \sigma_{in}$ can be made very close to 1, which the "ridge" positioned symmetrically as in Figure 8(C), for rather reasonable and easily experimentally achievable adjustments of the parameters. For example, if we had a system with a somewhat smaller reference logic-**1** value for the input hydrogen peroxide, $0.15 \mu M$ instead of the initially selected $0.25 \mu M$, then we would get the quality measure very close to 1, and a symmetrically positioned ridge, as can be discerned directly from Figure 11. (Note that the value of $\sigma_{out}^{max} / \sigma_{in}$ in Figure 11 "as is," without any parameter optimization, is close to 2.)

Recently, yet another not smooth-convex gate-function shape was realized. Electrode-immobilized enzyme glucose-6-phosphate dehydrogenase (G6PDH) was used[2] to catalyze an enzymatic reaction which carries out an **AND** logic gate with the response of the type shown in Figure 8(D) — sigmoid in one of the inputs. A kinetic model was developed[2] and utilized to evaluate the extent to which the experimentally realized gate was close to optimal, which in this case also corresponds to virtually no analog noise amplification.

However, there have been no experiments reported thus far that would realize the double-sigmoid gates, of the type of Figure 8(B), which actually suppress analog noise. Furthermore, even for one-sided-sigmoid gates, Figure 8(D), there has been no experimental utilization of those biocatalysts which have confirmed kinetic self-promoter properties with respect to one of the inputs, as is typical for many allosteric enzymes.[80] (In the above G6PDH case the origin of the self-promoter property is not fully understood[2] and could be electrode-immobilization induced.) Thus, a lot of experimental work is still needed to fully explore the richness of single-enzyme-based biocatalytic reactions for low-noise-amplification Boolean gate realizations.

## 6.    Networking Enzyme-Based Biochemical Reactions

As exemplified and discussed in Section 5, optimization of enzyme-based gates one at a time is in most cases not straightforward. Therefore, *relative* optimization of a *network* of enzymatic reactions as a whole has also been explored.[68] In this section, we illustrate this approach for an example of a network of three **AND** gates, shown in Figure 12. The coupled reactions are shown in Figure 12(A), and they include steps common in sensor development[81] for maltose and its sources. Figure 12(B) offers a modular network representation of the processes involved, in terms of three **AND** gates.

We note that this convenient representation is actually approximate to some extent, because it obscures some of the complexity of the processes involved. Specifically, β-amylase cuts β-maltose molecules off starch, whereas maltose phosphorylase takes in α-maltose. One of its products, in turn, is α-D-glucose, whereas the next enzyme, GDH, takes in β-D-glucose. Considering also that water is not really a variable input, a more precise, but not modular in terms of only **AND** gates, network representation is shown in Figure 12(C): The network actually involves two identity, **I**, functions, including one corresponding to the transduction of the output concentration to the optical readout signal. The two delayed-identity functions, **D**, correspond to β-maltose naturally converting[81,82] to α-maltose in aqueous environment, and α-glucose naturally converting[83] to β-glucose. These interconversion processes, involving equilibration between the α and β anomeric forms for each of the two saccharides, have time scales[84] of order 15 to 40 minutes. To speed up the dynamics, these experiments were carried out at $(50.0 \pm 0.1)^{\circ}$C; each data point was taken at $t^{\text{gate}} = 1000$ sec. Additional experimental details can be found in Ref. 68.

As mentioned in Section 5, we favor the use of a simple, few parameter description of the network elements that will allow us to "tweak" the relative gate activities in the network to improve its stability. For a one-variable convex function, variation of a gate's parameters rebalances the slopes near 0 and 1 with respect to each other: We need at least one phenomenological parameter to describe this. A convenient fit function is, for instance, $x(1+a)/(x+a)$. For a "generic" **AND** gate with response surface of the type of Figure 8(A), we use the product form,

$$F(x, y) = xy(1+a)(1+b)/(x+a)(y+b) \, , \qquad (6)$$

with two adjustable parameters, $0 < a, b \le \infty$ .

If this proposed approximate description is accurate for a given gate, then the parameters $a(E(0), t^{\text{gate}}, ...; r, R, ...; ...)$ and $b(E(0), t^{\text{gate}}, ...; r, R, ...; ...)$ will be functions of the directly adjustable variables such as concentrations of the enzyme, $E(0)$, and of other chemicals, the reaction time, $t^{\text{gate}}$, etc., as well as functions of parameters adjustable via changes in the physical or chemical conditions, such as the various process rates, here denoted by $r$, $R$, ... . Without detailed rate-equation kinetic modeling, this parameter dependence is not known, and we cannot guarantee that this functional form provides a good approximation for the actual $(x, y)$-dependence of the gate-function response surface. We will not attempt such a modeling for each gate in the network. Instead, we use the approximate parameterization, Eq. (6), to derive information on the relative network functioning by selective probes of responses to inputs, and we attempt to optimize the overall network performance.

The largest of the four gradient values of the fitting function, Eq. (6), is minimized at $a_{\text{optimal}} = b_{\text{optimal}} = 1/(\sqrt[4]{2} - 1) \approx 5.3$ . We also define

$$A = a/(1+a) \, , \qquad B = b/(1+b) \, , \qquad (7)$$

with optimal values $A_{\text{optimal}} = B_{\text{optimal}} = 2^{-1/4} \approx 0.84$ . The range for these parameters is $0 < A, B \le 1$ . We note that the actual gradient values in the symmetric $(a = b)$ case are $\sqrt{2} a/(1+a) = \sqrt{2} A$ for the logic-**11**, $1/A$ $(= 1/B)$ for **01** (**10**), and 0 for **00**. With the optimal parameter selections, the gradients at the three logic points **01**, **10**, **11**, are $\sqrt[4]{2} \approx 1.189$ . This means that the optimized gate-functions still somewhat amplify analog noise, by adding ~19% per processing step. This property of the convex gate response functions was mentioned in Section 5.

For the network in Figure 12(B), we follow the convention of numbering the gates in their sequence, counting from the output. Suppose that we set the inputs $x_{2,3} = 1$, and measure the function $z(x_1)$. Then it follows from Eq. (6) that the data should be fitted to the functional form $z(x_1) = x_1 /[(1 - A_1)x_1 + A_1]$, where for brevity we omitted all the fixed arguments. For variation of the output as a function of the other two inputs, we have $z(x_2) = x_2 /[(1 - A_2 B_1)x_2 + A_2 B_1]$ and

$$z(x_3) = x_3 /[(1 - A_3 B_2 B_1)x_3 + A_3 B_2 B_1].\qquad(8)$$

In terms of the variables $A$ and $B$ defined in Eq. (7), not only the input $x_1$ dependence, but also the input $x_{2,3}$ dependences require single-parameter fits. Thus, variation of $x_1$ provides information on one of the phenomenological parameters, $a_1$, of gate 1. Variation of $x_{2,3}$ does not actually lead to a more complicated several-parameter data fit, even though the signal being varied, goes through more than one gate before affecting the output. Instead, we get information on a combination of parameters from more than one gate.

The initial sets of data, one of which, for Eq. (8), is exemplified in Figure 13 for the fixed gate-time output signal definition (data were also gathered for the steady-sate rate-of-output signal definition), were collected with the experimentally convenient but otherwise initially randomly selected values for the adjustable "gate machinery" parameters[68] which we will limit here to the initial enzyme concentrations, $E_1(0)$, $E_2(0)$, $E_3(0)$, for definiteness. The data were rescaled into the logic variable ranges and fitted according to the above single-parameter equations. We note that since not all the inputs of all the gates are varied to probe the response of the final output, we do not get all the 6 phenomenological fit parameters. We only get one parameter and two additional combinations of parameters. Thus, we can only draw a limited set of conclusions regarding the network functioning. For the following discussion, the data were recast, see Table 1, in terms of the geometric means of the parameters that are known only as combinations (as products). Furthermore, since in the optimal-value case the gradients at the three non-**00** logic-points are actually $1/A$ (or $1/B$) and $\sqrt{2}A$ (or $\sqrt{2}B$), we took the maxima of these quantities to compare with the optimal (the smallest possible) value of the gradients, $\sqrt[4]{2} \approx 1.19$.

The following semi-quantitative conclusions follow from considering the initial data in Table 1. Both the time-dependence-slope based and the value (at $t^{\text{gate}}$) based signal definitions[68] give qualitatively similar results. Gate 3 seems to be the least noisy, whereas the larger "noise amplification measure" values that involve the other two gates should be attributed to gate 1 which contributes to both measures and is thus the primary candidate for parameter modification. In fact, for the initial experiment the maximal values in Table 1 were *all* realized with the $1/A$ or $1/B$ type value combinations, rather than the combinations involving $\sqrt{2}A(=\sqrt{2}a/(1+a))$ or $\sqrt{2}B$ type expressions. This suggests that the gradients are generally larger at logic-**01** points and **10** points, as compared to logic-**11** points. One way to decrease noise amplification in our network is thus to "shift" the gradients from lower to higher input concentrations. Larger variation of the output at large input values, will occur if we work less close to saturation, i.e., decrease the rates of (some of) the reactions.

Since gate 1 was already identified as candidate for adjustment, we selected to decrease the (initial) amount of the enzyme GDH. In fact, from arguments in Section 5 we expect that a significant reduction is needed to achieve a noticeable effect. A new set of data was measured,[68] with the concentration of GDH reduced by an order of magnitude (from 2.00 units to 0.18 units), as illustrated in Figure 13 for one of the inputs. The results of these "optimized" data fits are also summarized in Table 1. As already emphasized, we are aiming at identifying the regime of *reduced noise amplification*. From this point of view, the results are quite promising: With the use of our simple phenomenological data fitting functions, the noise-amplification measures (see Table 1) came out consistently lower (closer to the optimal) for the modified (improved) network as compared to the original one. Inspection of Figure 13 confirms out expectation that the data fits by the phenomenological two-parameter functional forms are at best semi-quantitative.

However, as far as avoiding noise amplification, the numbers in Table 1 are quite encouraging and support the validity of such "global" approaches to network modeling.

## 7.      Interfacing of Enzyme Logic with Signal Responsive Materials

Signals generated by enzyme logic systems in the form of concentration changes of reacting species can be read out using various analytical techniques: optical[57-59] and electrochemical.[2,75] Sensitive analytical methods and instruments are required because the concentration changes produced upon biocatalytic reactions occur usually at rather low levels. Application of highly sensitive techniques requires electronic transduction and amplification of the output signals produced by enzyme logic systems. This approach was actually also applied to most of chemical computing systems based on non-biological molecules.[4] Chemical changes in enzymatic biocatalytic systems can be coupled to signal-responsive materials resulting in changes of their bulk properties, thus amplifying the output signals generated by enzymes. Signal-responsive materials coupled with enzyme logic systems could be realized by polymers responding to external chemical signals by restructuring between swollen and shrunken states.[85] The structural reorganization of polymers will substantially amplify the chemical changes generated by the enzyme reactions, thus excluding the need for highly sensitive analytical techniques to observe the output signals from the biocomputing systems. Application by signal-responsive polymers in a biochemical environment already has a well established background, thus allowing their integration with biocomputing systems.[86]

This approach has lead to fabrication of "smart" multi-signal-responsive materials equipped with built-in Boolean logic.[87] Generally, such systems should be capable of switching physical properties (such as optical, electrical, magnetic, wettability, permeability, etc.) upon application of certain input chemical signals and according to the build-in logic program. Chemical reactions biocatalyzed by enzymes thus respond to the input chemical signals according to the Boolean logic and transfer the output signal to the responsive polymeric support. The chemical coupling between the enzymatic systems and polymeric supports can be based on electron or proton exchange between them. Electron exchange between the redox-active polymeric support and enzymatic system results in the reduction or oxidation of the redox polymer. Proton exchange between the polymeric support (polyelectrolyte in this case) and enzyme system yields different ionic states (protonated or deprotonated) of the polyelectrolyte support. The expected transformations of the polymeric support upon transition between different oxidation or protonation states result in changes of the composition/structure of the polymeric matrix. This in turn causes corresponding changes of the matrix-specific properties to be considered as the final output signals, see Figure 14. It should be noted that two distinct parts of the integrated systems will be responsible for the process: (*i*) The biochemical (enzymatic) systems responsive to the external chemical input signals will provide the variation of the system composition/structure according to the build-in Boolean logic. (*ii*) The polymeric supports will transduce the composition/structure changes to changes of the matrix properties, thus providing the transduction of the input chemical signals (addition of chemicals) into output physical signals (change of the optical, electrical, magnetic properties, change of wettability, permeability, etc., of the matrix). It should be emphasized that this transduction process follows the logic operations provided by the biochemical systems. Macroscopic changes in polymeric structure induced by enzyme reactions can eventually be used to design chemical actuators controlled by the signals processed by the enzyme logic systems.

Recent studies have shown that chemical transformations occurring at various interfaces and in polymeric matrices (induced by photochemical,[88] electrochemical,[89] magnetic,[90] or chemical/biochemical means[91]) can result in substantial changes in the properties of the materials (optical density, reflectivity, electrical conductivity, porosity/permeability, density/volume, wettability, etc.). Mixed-polymer systems (specifically polymer "brushes") were shown to be highly efficient responsive systems substantially changing their physical properties upon re-configuration of the components included in the mixed system.[92] Functional integration of enzymes, which operate as logic gates,[57,58] with the polymeric matrices allows Boolean treatment of the input chemical signals and the respective changes in the material properties. For example, two different input signals coming to the system will change the material properties according to the Boolean treatment of the input signals, mediated by electronic or ionic coupling between the enzymes and responsive polymer (exchange of electrons or protons resulting in the alteration of the oxidation or protonation state of the polymer), see Figure 14. It should be noted that the changes schematically shown in Figure 14 will only occur when the output signal of the enzyme logic gate is "TRUE" (**1**), while the "FALSE" (**0**) output will result in no changes in the system. If the system is changed (the "TRUE" output signal), it might be reset to the original state by chemical or electrochemical means (addition of chemicals to the solution or application of a potential on the conductive solid support). Then, the system will be ready to respond to the next set of input signals that may (or may not) change the polymer state according to the Boolean treatment of the new inputs.

Since many polymer systems are pH-sensitive and switchable, enzyme-based logic gates have been designed using enzymes as biocatalytic input signals, processing information according to the Boolean functions **AND / OR**, and generating pH changes as the outputs of the gates.[93–95] The **AND** gate performed a sequence of biocatalytic reactions, Figure 15(A): sucrose hydrolysis was biocatalyzed by invertase (Inv) yielding glucose, which was then oxidized by oxygen in the presence of glucose oxidase (GOx). The later reaction resulted in the formation of gluconic acid and therefore lowered the pH value of the solution. The absence of the enzymes was considered as the input signals **0**, while the presence of them at the experimentally convenient concentrations was interpreted as the input signals **1**. The biocatalytic reaction chain was activated only in the presence of both enzymes (Inv and GOx: input signals **1,1**) resulting in the decrease of the solution pH value, Figure 15(B). The absence of any of the two enzymes (input signals **0,1** or **1,0**) or both of them (input signals **0,0**) resulted in the inhibition of the gluconic acid formation and thus no pH changes were produced. This biocatalytic chain mimics the **AND** logic operation expressed by the standard truth table, Figure 15(C). After completion of the biocatalytic reactions and reaching the final pH value, the system might be reset to the original pH by using another biochemical reaction in the presence of urease and urea resulting in the production of ammonia and elevating pH, Figure 15(A). The performance of the biochemical system can be described in terms a logic circuitry with **AND/Reset** function, Figure 15(D).

Another gate, operating as Boolean **OR** logic function was composed of two parallel reactions, Figure 15(E): hydrolysis of ethyl butyrate and oxidation of glucose biocatalyzed by esterase (Est) and glucose oxidase (GOx), respectively, and resulting in the formation of butyric acid and gluconic acid. Any of the produced acids and both of them together resulted in the formation of acidic solution, Figure 15(F). Thus, in the absence of both enzymes (Est and GOx: input signals **0,0**) the two reactions were inhibited and the pH value was unchanged. When either enzyme (Est or GOx: input signals **0,1** or **1,0**) or both of them together (input signals **1,1**) were present, one or both of the reactions proceeded and resulted in the acidification of the solution. The features of the system correspond to the **OR** logic operation and can be expressed by the standard truth table, Figure 15(G). Similarly to the preceding example, the system can be reset to the initial pH by the urease catalyzed reaction allowing sequence of **OR** / **Reset** functions, Figure 15(H).

Enzyme-logic systems producing pH changes were coupled with various pH-sensitive polymer-functionalized nanostructured systems: membranes,[93] nanoparticle suspensions[74,94] and water/oil emulsions.[95] For example, the pH changes produced by the **AND** / **OR** enzyme logic gates shown in Figure 15 were coupled with a pH-responsive membrane, resulting in the opening/closing of the membrane pores, thus transducing the biochemical logic operation into the bulk material property change,[93] see Figure 16. The membrane was prepared by salt-induced phase separation of sodium alginate and gelatin and cross-linked by $CaCl_2$, Figure 16(A). The membrane operates by swelling its gel body in response to changes in pH. This leads to shrinkage of the pores and consequently to a change in its permeability. The membrane was deposited onto an ITO-glass electrode for electrochemical characterization or onto a porous substrate (track-etched polycarbonate membrane) for permeability measurements. The Scanning Probe Microscopy (SPM) topography images obtained *in situ* in a liquid cell, Figure 16(B), electrochemical impedance spectroscopy, Figure 16(C), and probe molecules (fluorescent dye Rhodamine B) diffusivity through the polyelectrolyte membrane, Figure 16(D), were all utilized to monitor the behavior of the membrane coupled with the enzyme-based logic gates. Strong dependence of the swelling of the polyelectrolyte membrane on pH was found: the pores are open at pH < 4 and completely closed at pH > 5. The pH changes were induced *in situ* using the enzyme logic gates shown in Figure 15. The corresponding SPM images, impedance and permeability changes were characteristic of the **AND** / **OR** logic gates, and after completion they were followed by the reset generated in the presence of urea and urease as described above.

Similarly to the logically controlled switchable membrane, other signal-responsive nanostructured materials were functionally coupled with enzyme logic systems. Reversible aggregation-dissociation of polymer-functionalized nanoparticles[94] and inversion of water-oil emulsion stabilized with nanoparticles[95] were controlled by the enzyme signals processed according to the **AND** / **OR** Boolean logic outlined in Figure 15. All the studied signal-responsive systems demonstrated Boolean logic operations encoded in the biochemical systems.

## 8. Interfacing of Enzyme Logic with Switchable Electrodes and Bioelectronic Devices

pH-switchable materials immobilized on interfaces of electronic/electrochemical transduces, e.g., Si-chips[96] or conducting electrodes,[97-100] were coupled with enzyme logic systems producing pH changes in solutions as logic responses to input signals. This allowed electronic transduction of the generated output signals, converting the systems into multi-signal biosensors chemically processing various patterns of the input signals using logic "programs" built-in in the enzyme systems. For example, enzyme logic systems mimicking Boolean **AND** / **OR** logic operations and producing the output signal in the form of solution pH changes were coupled with charging-discharging organic shells around Au nanoparticles associated with a Si-chip surface, Figure 17(A).[96] This resulted in the capacitance changes at the modified interface allowing electronic transduction of the biochemical signals processed by the enzyme logic systems, Figure 17(B-C).

Another approach to electrochemical transduction of the output signals generated by enzyme logic systems in the form of pH changes has been based on the application of polyelectrolyte-modified electrode surfaces.[97-100] Polyelectrolytes covalently bound to the electrode surface as polymer brushes offer pH-sensitivity allowing control of the electrode interfacial properties by varying pH values. Charged states of polymer brushes produce hydrophilic swollen thin-films at the electrode surface resulting in high permeability for transport of soluble redox probes to the conducting supports, thus yielding the electrochemically active state of the modified electrode. Upon discharging the polymer chains, the resulting hydrophobic shrunken brush isolated the conducting supports yielding the inactive state of the modified electrode. Switching between the ON and OFF states of the electrode modified with the polymer brush was achieved by varying the pH value of the solution. This property of the polymer brush-functionalized electrodes was used to couple them with an enzyme logic system composed of several networked gates.[98] The logic network composed of three enzymes (alcohol dehydrogenase, glucose dehydrogenase and glucose oxidase) operating in concert as four concatenated logic gates (**AND / OR**), was designed to process four different chemical input signals (NADH, acetaldehyde, glucose and oxygen), see Figure 18. The cascade of biochemical reactions resulted in pH changes controlled by the pattern of the applied biochemical input signals. The "successful" set of the inputs produced gluconic acid as the final product and yielded an acidic medium, lowering the pH of the solution from its initial value of pH 6-7 to the final value of ca. 4, thus switching ON the interface for the redox process of a diffusional redox probe, $[Fe(CN)_6]^{3-/4-}$. The chemical signals processed by the enzyme logic system and transduced by the sensing interface were read out by electrochemical means using cyclic voltammetry, see Figure 19(A). Reversible activation-inactivation of the electrochemical interface was achieved upon logic processing of the biochemical input signals and then by the reset function activated in the presence of urease and urea, Figure 19(A)-inset. The whole set of the input signal combinations included 16 variants, while only **0,0,1,1**; **0,1,1,1**; **1,0,1,1**; **1,1,1,0** and **1,1,1,1** combinations resulted in the ON state of the electrochemical interface, Figure 19(B). The present system exemplifies a multi-gate / multi-signal processing enzyme logic system associated with electrochemical transduction readout of the output signal.

Another pH-sensitive polymer-brush loaded with an electron transfer mediator $(Os(dmo-bpy)_2$, where dmo-bpy = 4,4'-dimethoxy-2,2'-bipyridine),[101] was applied to switch ON-OFF bioelectrocatalytic reactions upon receiving pH-signals from enzyme logic gates processing various biochemical inputs. In this way, electrocatalytic oxidation of NADH[97] and bioelectrocatalytic oxidation of glucose[100] were controlled by enzyme logic gates coupled with the switchable electrocatalytic interface through pH changes. This electrode was also integrated[102,103] into an enzyme-based biofuel cell serving there as a switchable biocatalytic cathode for oxygen reduction, see Figure 20. Glucose oxidation in the presence of soluble glucose oxidase (GOx) and methylene blue (MB) mediating electron transport to a bare ITO electrode was used as an anodic process. The biofuel cell was switched ON on-demand upon processing biochemical signals by enzyme logic gates and reset back to the OFF state by another biocatalytic reaction in the presence of urea and urease. The coupling of the *in situ* enzyme reactions with the switchable electrochemical interface was achieved by pH changes generated in the course of the enzymatic reactions resulting in swelling/shrinking the redox polymer at the electrode surface as described above. Early on,[102] single logic gates **AND / OR** were used to control the biofuel cell power production. Recently, a higher complexity biocatalytic system based on the concerted operation of four enzymes activated by four chemical input signals was designed, Figure 21(A), to mimic a logic network composed[103] of three logic gates: **AND / OR** connected in parallel and generating two intermediate signals for the final **AND** gate, see Figure 21(B). The switchable biofuel cell was characterized by measuring polarization curves at its "mute" and active states. A low voltage-current production was characteristic of the initial inactive state of the biofuel cell at pH ca. 6, Figure 22(A-a). Upon receiving an output signal in the form of a pH decrease from the enzyme logic network, the voltage-current production by the biofuel cell was dramatically enhanced when pH reached ca. 4.3, Figure 22(A-b). When activation of the biofuel cell was achieved, another biochemical signal (urea in the presence of urease) resulted in increase of pH, thus resetting the cell to its inactive state with the small voltage-current production, Figure 22(A-c). This cyclic operation of the biofuel cell in response to receiving biochemical signals can be followed by reversible changes of the current production, Figure 22(A)-inset. The biofuel cell switching from the "mute" state with low activity to the active state was achieved by the appropriate combination of the input signals processed by the enzyme logic network. Only three combinations of the input signals: **1,1,1,0**; **1,1,0,1** and **1,1,1,1** from the 16 possible variants resulted in the solution pH change, thus switching the biofuel cell to its active state, Figure 22(B). The studied biofuel cells exemplify a new type of bioelectronic devices with the bioelectronic function controlled by a biocomputing system. Such devices will provide a new dimension in bioelectronics and biocomputing benefiting from the integration of both concepts.

## 9.      Conclusions and Future Challenges

In summary, we reviewed various aspects of enzyme-based logic gates, their networking, and their interfacing with systems that offer transduction of the biochemical output into electronic or other signals. Let us now briefly comment on future challenges for this field of research.

We point out that an important development in enzyme-based logic has been initiated[3,104] with the introduction of the novel sensor concepts, with multiple input signals processed biochemically before transduction to the output. Such systems would require utilization of realistic, rather than model, biochemical inputs, with proper ranges for the "digital" **0** and **1** signal cutoffs, as well as "gate machinery" compatible with the environment of the sensing application. Furthermore, new demands for the biochemical gate networking will require a strategy to systematize the modeling,[3] utilizing modular network analysis and detailed network optimization approaches, supplemented with single-gate optimization of key elements in the network. As the biocomputing networks become larger and more complex, new gates and non-Boolean network elements will have to be realized, characterized, and modeled. The latter, non-Boolean elements should include filters, functioning, for instance, by diverting some of the output or input(s): It has been argued[67] that this can induce sigmoid behavior. More long-term projects will also address issues of signal slitting/balancing, and signal amplification. Ultimately, for larger networks the issues of network design for digital error correction will also come into play. Enzyme properties not presently explored for information processing, might offer interesting avenues for optimization. As already mentioned, allosteric enzymes[80] frequently have the self-promoter (sigmoid-response) property with respect to their substrate concentrations. Combination of enzymes with other biomolecules, specifically, with immune-recognition biomolecules (antigens–antibodies), offers interesting avenues for new information processing designs and applications. Preliminary work has already demonstrated logic operations performed by antigen–antibody interactions coupled with enzyme logic gates.[105] The logically processed immune-signals were utilized to control the operation of a biofuel cell.[106]

A variety of biomedical and biotechnological applications could be envisaged for the developed hybrid systems. These applications can range from biochemical processes (e.g., biomolecular oxidative damage)[107] to bioelectronic devices (e.g., biofuel cells)[102,103,106] controlled by enzyme logic systems. As mentioned, the biocomputing approach reviewed here, paves the way to novel digital biosensors processing multiple-biochemical signals and producing a combination of outputs dependent on a complex pattern of inputs. The biochemical signals can be processed by chemical means based on the enzyme logic systems and the difference between different physiological scenarios can be directly derived from the chemically processed information, hence obviating the need for computer analysis of the biosensing information. In addition to analysis of the data, the output signals might be directed to chemical actuators (e.g., signal responsive membranes) leading to on-demand drug release. A diverse range of "smart" (stimuli-responsive) materials, with switchable physical properties, has been developed for *in vivo* drug delivery.[108] The new hybrid materials with built-in Boolean logic will be capable of switching physical properties in response to the output of an enzyme-logic system towards autonomous on-demand drug delivery. The output signals generated by enzyme logic networks will activate "smart" chemical actuators, resulting, for example, in the opening of a membrane releasing a drug, could lead to a novel unconventional approach to the decision-making (sense-and-treat) biosensor. We anticipate that such devices controlled by biochemical logic networks will facilitate decision-making in connection to autonomous feedback-loop drug-delivery systems and will revolutionize the monitoring and treatment of patients, and many other biomedical applications.

The demonstrated approaches to interfacing of biomolecular computing systems with signal-responsive materials enable the use of various biocatalytic reactions to control the properties of responsive materials and systems with built-in Boolean logic. This approach would be an efficient way to fabricate "smart" multi-signal responsive drug delivery systems, sensors, miniaturized switchers, microfluidic devices, etc., which can function without communication with an external electronic computer revolutionizing many biotechnological applications.

## Acknowledgements

We gratefully acknowledge useful discussions, idea exchanges, and collaboration with our colleagues, Prof. L. Alfonta, L. Amir, M. Arugula, S. Chinnapareddy, Dr. L. Fedichkin, A. Gromenko, Dr. J. Halámek, Prof. A. Melman, Dr. D. Melnikov, Prof. S. Minko, Dr. V. Pedrosa, Dr. M. Pita, Prof. M. Privman, Prof. A. Simonian, Prof. I. Sokolov, Dr. D. Solenov, G. Strack, T. Tam, Prof. J. Wang, and J. Zhou, as well as support of our research programs by the NSF (Grants CCF-0726698 and DMR-0706209), by the Office of Naval Research (ONR Award #N00014-08-1-1202), and by the Semiconductor Research Corporation (Award 2008-RJ-1839G).

# REFERENCES

1. Unconventional Computation, *Lecture Notes in Computer Science*, Vol. **5715**, edited by C. S. Calude, J. F. Costa, N. Dershowitz, E. Freire and G. Rozenberg (Springer, Berlin, 2009).

2. V. Privman, V. Pedrosa, D. Melnikov, M. Pita, A. Simonian and E. Katz, *Biosens. Bioelectron.*, 2009, in press (available at http://dx.doi.org/10.1016/j.bios.2009.08.014).

3. E. Katz, V. Privman and J. Wang, submitted for publication (available at http://arxiv.org/abs/0909.1583).

4. (*a*) A. P. de Silva, S. Uchiyama, T. P. Vance and B. Wannalerse, *Coord. Chem. Rev.*, 2007, **251**, 1623-1632; (*b*) A. P. de Silva and S. Uchiyama, *Nature Nanotechnology*, 2007, **2**, 399-410; (*c*) K. Szacilowski, *Chem. Rev.*, 2008, **108**, 3481-3548.

5. (*a*) P. Fu, *Biotechnol. J.*, 2007, **2**, 91-101; (*b*) Y. Benenson, T. Paz-Elizur, R. Adar, E. Keinan, Z. Livneh and E. Shapiro, *Nature*, 2001, **414**, 430-434.

6. P. Dittrich, *Lect. Notes Computer Sci.*, 2005, **3566**, 19-32.

7. A. N. Shipway, E. Katz and I. Willner, in *Molecular Machines and Motors*, Structure and Bonding, Vol. **99**, Pages 237-281, edited by J.-P. Sauvage (Springer, Berlin, 2001).

8. (*a*) Y. L. Zhao, W. R. Dichtel, A. Trabolsi, S. Saha, I. Aprahamian and J. F. Stoddart, *J. Am. Chem. Soc.*, 2008, **130**, 11294-11295; (*b*) H. R. Tseng, S. A. Vignon, P. C. Celestre, J. Perkins, J. O. Jeppesen, A. Di Fabio, R. Ballardini, M. T. Gandolfi, M. Venturi, V. Balzani and J. F. Stoddart, *Chem. Eur. J.*, 2004, **10**, 155-172; (*c*) D. A. Weinberger, T. B. Higgins, C. A. Mirkin, C. L. Stern, L. M. Liable-Sands and A. L. Rheingold, *J. Am. Chem. Soc.*, 2001, **123**, 2503-2516.

9. (*a*) M. Suresh, A. Ghosh and A. Das, *Tetrahedron Lett.*, 2007, **48**, 8205-8208; (*b*) J. D. Crowley, D. A. Leigh, P. J. Lusby, R. T. McBurney, L. E. Perret-Aebi, C. Petzold, A. M. Z. Slawin and M. D. Symes, *J. Am. Chem. Soc.*, 2007, **129**, 15085-15090.

10. (*a*) S. Iwata and K. Tanaka, *J. Chem. Soc., Chem. Commun.*, 1995, 1491-1492; (*b*) S. H. Lee, J. Y. Kim, S. K. Kim, J. H. Leed and J. S. Kim, *Tetrahedron*, 2004, **60**, 5171-5176; (*c*) D. C. Magri, G. J. Brown, G. D. McClean and A. P. de Silva, *J. Am. Chem. Soc.*, 2006, **128**, 4950-4951; (*d*) K. K. Sadhu, B. Bag and P. K. Bharadwaj, *J. Photochem. Photobiol. A*, 2007, **185**, 231-238.

11.  (*a*) S. Giordani, M. A. Cejas and F. M. Raymo, *Tetrahedron*, 2004, **60**, 10973-10981; (*b*) E. Katz, A. N. Shipway and I. Willner, in *Photoreactive Organic Thin Films*, edited by S. Sekkat and W. Knoll, Chapter II-7, Pages 219-268 (Elsevier, San Diego, USA, 2002); (*c*) V. Balzani, *Photochem. Photobiol. Sci.*, 2003, **2**, 459-476.

12.  (*a*) F. M. Raymo and S. Giordani, *Proc. Natl. Acad. USA*, 2002, **99**, 4941-4944; (*b*) M. Lion-Dagan, E. Katz and I. Willner, *J. Am. Chem. Soc.*, 1994, **116**, 7913-7914; (*c*) I. Willner, M. Lion-Dagan, S. Marx-Tibbon and E. Katz, *J. Am. Chem. Soc.*, 1995, **117**, 6581-6592; (*d*) A. Doron, M. Portnoy, M. Lion-Dagan, E. Katz and I. Willner, *J. Am. Chem. Soc.*, 1996, **118**, 8937-8944; (*e*) N. G. Liu, D. R. Dunphy, P. Atanassov, S. D. Bunge, Z. Chen, G. P. Lopez, T. J. Boyle and C. J. Brinker, *Nano Lett.*, 2004, **4**, 551-554; (*f*) Z. F. Liu, K. Hashimoto and A. Fujishima, *Nature*, 1990, **347**, 658-660; (*g*) E. Katz and A. N. Shipway, in *Bioelectronics: From Theory to Applications*, edited by I. Willner and E. Katz, Chapter 11, Pages 309-338 (Wiley-VCH, Weinheim, 2005); (*h*) S. Bonnet and J. P. Collin, *Chem. Soc. Rev.*, 2008, **37**, 1207-1217; (*h*) P. R. Ashton, R. Ballardini, V. Balzani, A. Credi, K. R. Dress, E. Ishow, C. J. Kleverlaan, O. Kocian, J. A. Preece, N. Spencer, J. F. Stoddart, M. Venturi and S. Wenger, *Chem. Eur. J.*, 2000, **6**, 3558-3574; (*i*) I. Thanopulos, P. Kral, M. Shapiro and E. Paspalakis, *J. Modern Optics*, 2009, **56**, 1-18.

13.  (*a*) I. M. Hsing, Y. Xu and W. T. Zhao, *Electroanalysis*, 2007, **19**, 755-768; (*b*) E. Katz, R. Baron and I. Willner, *J. Am. Chem. Soc.*, 2005, **127**, 4060-4070; (*c*) E. Katz, L. Sheeney-Haj-Ichia, B. Basnar, I. Felner and I. Willner, *Langmuir*, 2004, **20**, 9714-9719; (*d*) I. Willner and E. Katz, *Angew. Chem. Int. Ed.*, 2003, **42**, 4576-4588; (*e*) E. Katz, L. Sheeney-Haj-Ichia and I. Willner, *Chem. Eur. J.*, 2002, **8**, 4138-4148; (*f*) R. Hirsch, E. Katz and I. Willner, *J. Am. Chem. Soc.*, 2000, **122**, 12053-12054; (*g*) J. Wang and A. N. Kawde, *Electrochem. Commun.*, 2002, **4**, 349-352; (*h*) R. Laocharoensuk, A. Bulbarello, S. Mannino and J. Wang, *Chem. Commun.*, 2007, 3362-3364; (*i*) J. Wang, *Electroanalysis*, 2008, **20**, 611-615; (*j*) O. A. Loaiza, R. Laocharoensuk, J. Burdick, M. C. Rodriguez, J. M. Pingarron, M. Pedrero and J. Wang, *Angew. Chem. Int. Ed.*, 2007, **46**, 1508-1511; (*k*) J. Wang, M. Scampicchio, R. Laocharoensuk, F. Valentini, O. Gonzalez-Garcia and J. Burdick, *J. Am. Chem. Soc.*, 2006, 128, 4562-4563; (*l*) J. Lee, D. Lee, E. Oh, J. Kim, Y. P. Kim, S. Jin, H. S. Kim, Y. Hwang, J. H. Kwak, J. G. Park, C. H. Shin, J. Kim and T. Hyeon, *Angew. Chem. Int. Ed.*, 2005, **44**, 7427-7432.

14. (*a*) L. Zheng and L. Xiong, *Colloids Surf. A*, 2006, **289**, 179-184; (*b*) M. Riskin, B. Basnar, E. Katz and I. Willner, *Chem. Eur. J.*, 2006, **12**, 8549-8557; (*c*) M. Riskin, B. Basnar, V. I. Chegel, E. Katz, I. Willner, F. Shi and X. Zhang, *J. Am. Chem. Soc.*, 2006, **128**, 1253-1260; (*d*) V. I. Chegel, O. A. Raitman, O. Lioubashevski, Y. Shirshov, E. Katz and I. Willner, *Adv. Mater.*, 2002, **14**, 1549-1553; (*e*) X. T. Le, P. Jégou, P. Viel, S. Palacin, *Electrochem. Commun.*, 2008, **10**, 699-703; (*f*) I. Poleschak, J. M. Kern and J. P. Sauvage, *Chem. Commun.*, 2004, 474-476.

15. (*a*) P. R. Ashton, R. Ballardini, V. Balzani, I. Baxter, A. Credi, M. C. T. Fyfe, M. T. Gandolfi, M. Gomez-Lopez, M. V. Martinez-Diaz, A. Piersanti, N. Spencer, J. F. Stoddart, M. Venturi, A. J. P. White and D. J. Williams, J. Am. Chem. Soc., 1998, **120**, 11932-11942; (*b*) C. J. Richmond, A. D. C. Parenty, Y. F. Song, G. Cooke and L. Cronin, *J. Am. Chem. Soc.*, 2008, **130**, 13059-13065; (*c*) F. Coutrot, C. Romuald and E. Busseron, *Org. Lett.*, 2008, **10**, 3741-3744; (*d*) Y. Shiraishi, Y. Tokitoh, G. Nishimura and T. Hirai, *Org. Lett.*, 2005, **7**, 2611-2614.

16. (*a*) G. Nishimura, K. Ishizumi, Y. Shiraishi and T. Hirai, *J. Phys. Chem. B* 2006, **110**, 21596-21602; (*b*) W. D. Zhou, J. B. Li, X. R. He, C. H. Li, J. Lv, Y. L. Li, S. Wang, H. B. Liu and D. B. Zhu, *Chem. Eur. J.* 2008, **14**, 754-763. (*c*) R. P. Fahlman, M. Hsing, C. S. Sporer-Tuhten and D. Sen, *Nano Lett.* 2003, **3**, 1073-1078; (*d*) Y. Shiraishi, Y. Tokitoh and T. Hirai, *Chem. Commun.* 2005, 5316-5318.

17. R. Baron, A. Onopriyenko, E. Katz, O. Lioubashevski, I. Willner, S. Wang and H. Tian, *Chem. Commun.*, 2006, 2147-2149.

18. A. Doron, M. Portnoy, M. Lion-Dagan, E. Katz and I. Willner, *J. Am. Chem. Soc.*, 1996, **118**, 8937-8944.

19. M. Biancardo, C. Bignozzi, H. Doyle and G. Redmond, *Chem. Commun.*, 2005, 3918-3920.

20. (*a*) B. Raychaudhuri and S. Bhattacharyya, *Appl. Phys. B*, 2008, **91**, 545-550; (*b*) S. Giordani and F. M. Raymo, *Org. Lett.*, 2003, **5**, 3559-3562.

21. (*a*) S. D. Straight, J. Andrasson, G. Kodis, S. Bandyopadhyay, R. H. Mitchell, T. A. Moore, A. L. Moore and D. Gust, *J. Am. Chem. Soc.*, 2005, **127**, 9403-9409; (*b*) C.-J. Fang, Z. Zhu, W. Sun, C.-H. Xu and C.-H. Yan, *New J. Chem.*, 2007, **31**, 580-586; (*c*) Z. Wang, G.

Zheng and P. Lu, *Organic Lett.*, 2005, **7**, 3669-3672; (*d*) S. D. Straight, P. A. Liddell, Y. Terazono, T. A. Moore, A. L. Moore and D. Gust, *Adv. Funct. Mater.*, 2007, **17**, 777-785.

22. (*a*) K. Szaciłowski, W. Macyk and G. Stochel, *J. Am. Chem. Soc.*, 2006, **128**, 4550-4551; (*b*) G. Wen, J. Yan, Y. Zhou, D. Zhang, L. Mao and D. Zhu, *Chem. Commun.*, 2006, 3016-3018; (*c*) F. Li, M. Shi, C. Huang and L. Jin, *J. Mater. Chem.*, 2005, **15**, 3015-3020.

23. (*a*) J. H. Qian, Y. F. Xu, X. H. Qian and S. Y. Zhang, *ChemPhysChem*, 2008, **9**, 1891-1898; (*b*) G. Fioravanti, N. Haraszkiewicz, E. R. Kay, S. M. Mendoza, C. Bruno, M. Marcaccio, P. G. Wiering, F. Paolucci, P. Rudolf, A. M. Brouwer and D. A. Leigh, *J. Am. Chem. Soc.*, 2008, **130**, 2593-2601.

24. (*a*) A. P. de Silva, *Nature*, 2008, **454**, 417-418; (*b*) P. M. Mendes, *Chem. Soc. Rev.*, 2008, **37**, 2512-2529; (*c*) S. Nitahara, N. Terasaki, T. Akiyama and S. Yamada, *Thin Solid Films*, 2006, **499**, 354-358; (*d*) T. Gupta and M. E. van der Boom, *Angew. Chem. Int. Ed.*, 2008, **47**, 5322-5326.

25. A. Credi, V. Balzani, S. J. Langford and J. F. Stoddart, *J. Am. Chem. Soc.*, 1997, **119**, 2679-2681.

26. (*a*) A. P. de Silva, H. Q. N. Gunaratne and C. P. McCoy, *Nature*, 1993, **364**, 42-44; (*b*) A. P. de Silva, H. Q. N. Gunaratne and C. P. McCoy, *J. Am. Chem. Soc.*, 1997, **119**, 7891-7892.

27. A. P. de Silva, H. Q. N. Gunaratne and G. E. M. Maguire, *J. Chem. Soc., Chem. Commun.*, 1994, 1213-1214.

28. A. P. de Silva and N. D. McClenaghan, *Chem. Eur. J.*, 2002, **8**, 4935-4945.

29. (*a*) A. P. de Silva, I. M. Dixon, H. Q. N. Gunaratne, T. Gunnlaugsson, P. R. S. Maxwell and T. E. Rice, *J. Am. Chem. Soc.*, 1999, **121**, 1393-1394; (*b*) S. D. Straight, P. A. Liddell, Y. Terazono, T. A. Moore, A. L. Moore and D. Gust, *Adv. Funct. Mater.*, 2007, **17**, 777-785; (*c*) B. Turfan and E. U. Akkaya, *Org. Lett.*, 2002, **4**, 2857-2859; (*d*) Z. Wang, G. Zheng and P. Lu, *Org. Lett.*, 2005, **7**, 3669-3672.

30. (*a*) H. T. Baytekin and E. U. Akkaya, *Org. Lett.*, 2000, **2**, 1725-1727; (*b*) G. Zong, L. Xiana and G. Lua, *Tetrahedron Lett.*, 2007, **48**, 3891-3894.

31. (*a*) T. Gunnlaugsson, D. A. Mac Dónaill and D. Parker, *J. Am. Chem. Soc.*, 2001, **123**, 12866-12876; (*b*) T. Gunnlaugsson, D. A. MacDónaill and D. Parker, *Chem. Commun.*, 2000, 93-94; (*c*) T. Gunnlaugsson, D. A. MacDonaill and D. Parker, *Chem. Commun.*, 2000, 93-94; (*d*) M. de Sousa, B. de Castro, S. Abad, M. A. Miranda and U. Pischel, *Chem. Commun.*, 2006, 2051-2053; (*e*) L. Li, M.-X. Yu, F. Y. Li, T. Yi and C. H. Huang, *Colloids Surf. A*, 2007, **304**, 49-53; (*f*) T. Gunnlaugsson, D. A. Mac Dónail and D. Parker, *Chem. Commun.*, 2000, 93-94.

32.  (*a*) V. Luxami and S. Kumar, *New J. Chem.*, 2008, **32**, 2074-2079; (*b*) J. H. Qian, X. H. Qian, Y. F. Xu and S. Y. Zhang, *Chem. Commun.*, 2008, 4141-4143.

33.  E. Pérez-Inestrosa, J.-M. Montenegro, D. Collado, R. Suau and J. Casado, *J. Phys. Chem. C*, 2007, **111**, 6904-6909.

34.  (a) W. Sun, C. H. Xu, Z. Zhu, C. J. Fang and C. H. Yan, *J. Phys. Chem. C*, 2008, **112**, 16973-16983; (*b*) Z. X. Li, L. Y. Liao, W. Sun, C. H. Xu, C. Zhang, C. J. Fang and C. H. Yan, *J. Phys. Chem. C*, 2008, **112**, 5190-5196; (*c*) A. Coskun, E. Deniz and E. U. Akkaya, *Org. Lett.*, 2005, **7**, 5187-5189; (*d*) D. Jiménez, R. Martínez-Máñez, F. Sancenón, J. V. Ros-Lis, J. Soto, A. Benito and E. García-Breijo, *Eur. J. Inorg. Chem.*, 2005, 2393-2403.

35.  (*a*) W. Sun, Y.-R. Zheng, C.-H. Xu, C.-J. Fang and C.-H. Yan, *J. Phys. Chem. C*, 2007, **111**, 11706-11711; (*b*) Y. Zhou, H. Wu, L. Qu, D. Zhang and D. Zhu, *J. Phys. Chem. B*, 2006, **110**, 15676-15679.

36.  U. Pischel and B. Heller, *New J. Chem.*, 2008, **32**, 395-400.

37.  (*a*) J. Andreasson, S. D. Straight, S. Bandyopadhyay, R. H. Mitchell, T. A. Moore, A. L. Moore and D. Gust, *J. Phys. Chem. C*, 2007, **111**, 14274-14278; (*b*) M. Amelia, M. Baroncini and A. Credi, *Angew. Chem. Int. Ed.*, 2008, **47**, 6240-6243; (*c*) E. Perez-Inestrosa, J. M. Montenegro, D. Collado and R. Suau, *Chem. Commun.*, 2008, 1085-1087.

38.  J. Andreasson, S. D. Straight, T. A. Moore, A. L. Moore and D. Gust, *J. Am. Chem. Soc.*, 2008, **130**, 11122-11128.

39.  (*a*) D. Margulies, C. E. Felder, G. Melman and A. Shanzer, *J. Am. Chem. Soc.*, 2007, **129**, 347-354; (*b*) M. Suresh, A. Ghosh and A. Das, *Chem. Commun.*, 2008, 3906-3908.

40.  (*a*) M. N. Chatterjee, E. R. Kay and D. A. Leigh, *J. Amer. Chem. Soc.*, 2006, **128**, 4058-4073; (*b*) R. Baron, A. Onopriyenko, E. Katz, O. Lioubashevski, I. Willner, S. Wang and H. Tian, *Chem. Commun.*, 2006, 2147-2149; (*c*) E. Katz, I. Willner, *Chem. Commun.*, 2005, 5641-5643; E. Katz and I. Willner, *Electrochem. Commun.*, 2006, **8**, 879-882; (*d*) F. Galindo, J. C. Lima, S. V. Luis, A. J. Parola and F. Pina, *Adv. Funct. Mater.*, 2005, **15**, 541-545; (*e*) A. Bandyopadhyay and A. J. Pal, *J. Phys. Chem. B*, 2005, **109**, 6084-6088; (*f*)

D. Fernandez, A. J. Parola, L. C. Branco, C. A. M. Afonso and F. Pina, *J. Photochem. Photobiol. A*, 2004, **168**, 185-189; (*g*) F. Pina, J. C. Lima, A. J. Parola and C. A. M. Afonso, *Angew. Chem. Int. Ed.*, 2004, **43**, 1525-1527; (*h*) G. Will, J. S. S. N. Rao and D. Fitzmaurice, *J. Mater. Chem.*, 1999, **9**, 2297-2299. (*i*) J. Hiller and M. F. Rubner, *Macromolecules*, 2003, **36**, 4078-4083; (*j*) F. Pina, A. Roque, M. J. Melo, I. Maestri, L. Belladelli and V. Balzani, *Chem. Eur. J.*, 1998, **4**, 1184-1191.

41.  R. Stadler, S. Ami, C. Joachim and M. Forshaw, *Nanotechnology*, 2004, **15**, S115-S121.

42.  A. P. De Silva, Y. Leydet, C. Lincheneau and N. D. McClenaghan, *J. Phys. Cond. Mat.*, 2006, **18**, S1847-S1872.

43.  A. Adamatzky, *IEICE Trans. Electron.*, 2004, **E87C**, 1748-1756.

44.  G. Bell and J. N. Gray, in: *Beyond Calculation: The Next Fifty Years of Computing*, edited by P. J. Denning and R. M. Metcalfe, Chapter 1, Pages 5-32 (Copernicus/Springer, NY, 1997).

45.  (*a*) K. P. Zauner, *Critical Rev. Solid State Mater. Sci.*, 2005, **30**, 33-69; (*b*) L. Fu, L. C. Cao, Y. Q. Liu and D. B. Zhu, *Adv. Colloid Interf. Sci.*, 2004, **111**, 133-157.

46.  (*a*) U. Pischel, *Angew. Chem. Int. Ed.*, 2007, **46**, 4026-4040; (*b*) G. J. Brown, A. P. de Silva and S. Pagliari, *Chem. Commun.*, 2002, 2461-2463.

47.  (*a*) D.-H. Qu, Q.-C. Wang and H. Tian, *Angew. Chem. Int. Ed.*, 2005, **44**, 5296-5299; (*b*) J. Andréasson, S. D. Straight, G. Kodis, C.-D. Park, M. Hambourger, M. Gervaldo, B. Albinsson, T. A. Moore, A. L. Moore and Devens Gust, *J. Am. Chem. Soc.*, 2006, **128**, 16259-16265; (*c*) J. Andréasson, G. Kodis, Y. Terazono, P. A. Liddell, S. Bandyopadhyay, R. H. Mitchell, T. A. Moore, A. L. Moore and D. Gust, *J. Am. Chem. Soc.*, 2004, **126**, 15926-15927; (*d*) M. V. Lopez, M. E. Vazquez, C. Gomez-Reino, R. Pedrido and M. R. Bermejo, *New J. Chem.*, 2008, **32**, 1473-1477.

48.  (*a*) D. Margulies, G. Melman and A. Shanzer, *J. Am. Chem. Soc.*, 2006, **128**, 4865-4871; (*b*) O. Kuznetz, H. Salman, N. Shakkour, Y. Eichen and S. Speiser, *Chem. Phys. Lett.*, 2008, **451**, 63-67.

49.  (*a*) Y. Liu, W. Jiang, H.-Y. Zhang and C.-J. Li, *J. Phys. Chem. B*, 2006, **110**, 14231-14235; (*b*) X. Guo, D. Zhang, G. Zhang and D. Zhu, *J. Phys. Chem. B*, 2004, **108**, 11942-11945; (*c*) F. M. Raymo and S. Giordani, *J. Am. Chem. Soc.*, 2001, **123**, 4651-4652.

50. F. Pina, M. J. Melo, M. Maestri, P. Passaniti and V. Balzani, *J. Am. Chem. Soc.*, 2000, **122**, 4496-4498.

51. A. H. Flood, R. J. A. Ramirez, W. Q. Deng, R. P. Muller, W. A. Goddard and J. F. Stoddart, *Austr. J. Chem.*, 2004, **57**, 301-322.

52. (*a*) X. G. Shao, H. Y. Jiang, W. S. Cai, *Prog. Chem.* 2002, **14**, 37-46; (*b*) A. Saghatelian, N. H. Volcker, K. M. Guckian, V. S. Y. Lin, M. R. Ghadiri, *J. Am. Chem. Soc.* 2003, **125**, 346-347; (*c*) G. Ashkenasy, M. R. Ghadiri, *J. Am. Chem. Soc.* 2004, **126**, 11140-11141.

53. (*a*) S. Sivan and N. Lotan, *Biotechnol. Prog.*, 1999, **15**, 964-970; (*b*) S. Sivan, S. Tuchman and N. Lotan, *Biosystems*, 2003, **70**, 21-33; (*c*) A. S. Deonarine, S. M. Clark and L. Konermann, *Future Generation Computer Systems*, 2003, **19**, 87-97; (*d*) G. Ashkenazi, D. R. Ripoll, N. Lotan and H. A. Scheraga, *Biosens. Bioelectron.*, 1997, **12**, 85-95; (*e*) R. Unger and J. Moult, *Proteins*, 2006, **63**, 53-64.

54. (*a*) M. N. Stojanovic, D. Stefanovic, T. LaBean and H. Yan, in *Bioelectronics: From Theory to Applications*, edited by I. Willner and E. Katz, Chapter 14, Pages 427-455 (Wiley-VCH, Weinheim, 2005); (*b*) A. Saghatelian, N. H. Volcker, K. M. Guckian, V. S. Y. Lin and M. R. Ghadiri, *J. Am. Chem. Soc.*, 2003, **125**, 346-347; (*c*) G. Ashkenasy and M. R. Ghadiri, *J. Am. Chem. Soc.*, 2004, **126**, 11140-11141.

55. M. N. Win and C. D. Smolke, *Science*, 2008, **322**, 456-460.

56. M. L. Simpson, G. S. Sayler, J. T. Fleming and B. Applegate, *Trends Biotechnol.*, 2001, **19**, 317-323.

57. G. Strack, M. Pita, M. Ornatska and E. Katz, *ChemBioChem*, 2008, **9**, 1260-1266.

58. R. Baron, O. Lioubashevski, E. Katz, T. Niazov and I. Willner, *J. Phys. Chem. A*, 2006, **110**, 8548-8553.

59. R. Baron, O. Lioubashevski, E. Katz, T. Niazov and I. Willner, *Angew. Chem. Int. Ed.*, 2006, **45**, 1572-1576.

60. T. Niazov, R. Baron, E. Katz, O. Lioubashevski and I. Willner, *Proc. Natl. Acad. USA*, 2006, **103**, 17160-17163.

61. G. Strack, M. Ornatska, M. Pita and E. Katz, *J. Am. Chem. Soc.*, 2008, **130**, 4234-4235.

62. (*a*) J. Xu and G. J. Tan, *J. Comput. Theor. Nanosci.*, 2007, **4**, 1219-1230; (*b*) M. Soreni, S. Yogev, E. Kossoy, Y. Shoham and E. Keinan, *J. Amer. Chem. Soc.*, 2005, **127**, 3935-3943.

63. M. Kahan, B. Gil, R. Adar and E. Shapiro, *Physica D*, 2008, **237**, 1165-1172.

64. U. Alon, *An Introduction to Systems Biology. Design Principles of Biological Circuits* (Chapman & Hall/CRC Press, Boca Raton, Florida, 2007).

65. *Quantum Information Science and Technology Roadmapping Project*, maintained online at http://qist.lanl.gov by the U.S. Department of Energy.

66. (*a*) J. Wakerly, Digital Design: Principles and Practices (Pearson/Prentice Hall, Upper Saddle River, NJ, 2005); (*b*) N. H. E. Weste and D. Harris, CMOS VLSI Design: A Circuits and Systems Perspective (Pearson/Addison-Wesley, Boston, 2004).

67. V. Privman, G. Strack, D. Solenov, M. Pita and E. Katz, *J. Phys. Chem. B*, 2008, **112**, 11777-11784.

68. V. Privman, M. A. Arugula, J. Halamek, M. Pita and E. Katz, *J. Phys. Chem. B*, 2009, **113**, 5301-5310.

69. M. A. Arugula, J. Halamek, E. Katz, D. Melnikov, M. Pita, V. Privman and G. Strack, in *Proc. Conf. CENICS 2009*, in press (IEEE Comp. Soc. Conf. Publ. Serv., Los Alamitos, California, 2009).

70. (*a*) Y. Setty, A. E. Mayo, M. G. Surette and U. Alon, *Proc. Natl. Acad. USA*, 2003, **100**, 7702-7707; (*b*) N. E. Buchler, U. Gerland and T. Hwa, *Proc. Natl. Acad. USA*, 2005, **102**, 9559-9564.

71. (*a*) O. Pretzel, *Error-Correcting Codes and Finite Fields* (Oxford Univ. Press, Oxford, 1992); (*b*) B. Arazi, *A Commonsense Approach to the Theory of Error Correcting Codes* (MIT Press, Cambridge, MA, 1988).

72. L. Fedichkin, E. Katz and V. Privman, *J. Comput. Theor. Nanosci.*, 2008, **5**, 36-43.

73. D. Melnikov, G. Strack, M. Pita, V. Privman and E. Katz, *J. Phys. Chem. B*, 2009, **113**, 10472-10479.

74. M. Pita, M. Krämer, J. Zhou, A. Poghossian, M. J. Schöning, V. M. Fernández and E. Katz, *ACS Nano*, 2008, **2**, 2160-2166.

75. M. Pita and E. Katz, *J. Am. Chem. Soc.*, 2008, **130**, 36-37.

76. R. E. Childs and W. G. Bardsley, *Biochem. J.*, 1975, **145**, 93-103.

77. (*a*) B. B. Hasinoff and H. B. Dunford, *Biochemistry*, 1970, **9**, 4930-4939; (*b*) A. T. Smith, S. A. Sanders, R. N. F. Thorneley, J. F. Burke and R. R. C. Bray, *Eur. J. Biochem*, 1992, **207**, 507-519.

78. D. V. Roberts, *Enzyme Kinetics* (Cambridge Univ. Press, Cambridge, UK, 1977).

79. (*a*) O. Feinerman, A. Rotem and E. Moses, *Nature Physics*, 2008, **4**, 967-973; (*b*) O. Feinerman and E. Moses, *J. Neurosci.*, 2005, **26**, 4526-4534.

80. J. Ricard and A. Cornish-Bowden, *Eur. J. Biochem.*, 1987, **166**, 255-272.

81. Y. Shirokane, K. Ichikawa and M. Suzuki, *Carbohydrate Res.*, 2000, **329**, 699-702.

82. (*a*) S. E. Weise, K. S. Kim, R. P. Stewart and T. D. Sharkey, *Plant Physiology*, 2005, **137**, 756-761; (*b*) Y. Tsumuraya, C. F. Brewer and E. J. Hehre, *Arch. Biochem. Biophys.*, 1990, **281**, 58-65.

83. D. L. Nelson and M. M. Cox, *Lehninger Principles of Biochemistry*, 4[th] Edition (W.H. Freeman and Company, New York, 2005), Page 242.

84. (*a*) Y. Shirokane and M. Suzuki, *FEBS Lett.*, 1995, **367**, 177-179; (*b*) M. Pagnotta, C. L. F. Pooley, B. Gurland and M. Choi, *J. Phys. Organic Chem.*, 1993, **6**, 407-411.

85. (*a*) I. Tokarev, S. Minko, *Soft Matter*, 2009, **5**, 511-524; (*b*) S. K. Ahn, R. M. Kasi, S. C. Kim, N. Sharma, Y. X. Zhou, *Soft Matter*, 2008, **4**, 1151-1157; (*c*) K. Glinel, C. Dejugnat, M. Prevot, B. Scholer, M. Schonhoff, R. V. Klitzing, *Colloids Surf. A*, 2007, **303**, 3-13.

86. P. M. Mendes, *Chem. Soc. Rev.*, 2008, **37**, 2512-2529.

87. M. Pita, S. Minko, E. Katz, *J. Mater. Sci.: Materials in Medicine*, 2009, **20**, 457-462.

88. I. Willner, A. Doron and E. Katz, *J. Phys. Org. Chem.*, 1998, **11**, 546-560.

89. V. I. Chegel, O. A. Raitman, O. Lioubashevski, Y. Shirshov, E. Katz and I. Willner, *Adv. Mater.*, 2002, **14**, 1549-1553.

90. E. Katz, L. Sheeney-Haj-Ichia, B. Basnar, I. Felner and I. Willner, *Langmuir*, 2004, **20**, 9714-9719.

91. X. Wang, Z. Gershman, A. B. Kharitonov, E. Katz and I. Willner, *Langmuir*, 2003, **19**, 5413-5420.

92. (*a*) I. Luzinov, S. Minko and V. V. Tsukruk, *Prog. Polym. Sci.*, 2004, **29**, 635-698; (*b*) S. Minko, *Polymer Rev.*, 2006, **46**, 397-420.

93. I. Tokarev, V. Gopishetty, J. Zhou, M. Pita, M. Motornov, E. Katz and S. Minko, *ACS Appl. Mater. Interfaces*, 2009, **1**, 532-536.

94. M. Motornov, J. Zhou, M. Pita, V. Gopishetty, I. Tokarev, E. Katz and S. Minko, *Nano Lett.*, 2008, **8**, 2993-2997.

95. M. Motornov, J. Zhou, M. Pita, I. Tokarev, V. Gopishetty, E. Katz and S. Minko, *Small*, 2009, **5**, 817-820.

96. M. Krämer, M. Pita, J. Zhou, M. Ornatska, A. Poghossian, M. J. Schöning and E. Katz, *J. Phys. Chem. B*, 2009, **113**, 2573-2579.

97. J. Zhou, T. K. Tam, M. Pita, M. Ornatska, S. Minko and E. Katz, *ACS Appl. Mater. Interfaces*, 2009, **1**, 144-149.

98. M. Privman, T. K. Tam, M. Pita and E. Katz, *J. Am. Chem. Soc.*, 2009, **131**, 1314-1321.

99. X. Wang, J. Zhou, T. K. Tam, E. Katz and M. Pita, *Bioelectrochemistry*, 2009, **77**, 69-73.

100. T. K. Tam, J. Zhou, M. Pita, M. Ornatska, S. Minko and E. Katz, *J. Am. Chem. Soc.*, 2008, **130**, 10888-10889.

101. T. K. Tam, M. Ornatska, M. Pita, S. Minko and E. Katz, *J. Phys. Chem. C*, 2008, **112**, 8438-8445.

102. L. Amir, T. K. Tam, M. Pita, M. M. Meijler, L. Alfonta and E. Katz, *J. Am. Chem. Soc.*, 2009, **131**, 826-832.

103. T. K. Tam, M. Pita, M. Ornatska and E. Katz, *Bioelectrochemistry*, 2009, **76**, 4-9.

104. (a) M. Pita, J. Zhou, K. M. Manesh, J. Halámek, E. Katz and J. Wang, *Sens. Actuat. B*, 2009, **139**, 631-636; (b) K. M. Manesh, J. Halámek, M. Pita, J. Zhou, T. K. Tam, P. Santhosh, M.-C. Chuang, J. R. Windmiller, D. Abidin, E. Katz and J. Wang, *Biosens. Bioelectron.*, 2009, **24**, 3569-3574.

105. G. Strack, S. Chinnapareddy, D. Volkov, J. Halámek, M. Pita, I. Sokolov and E. Katz**,** *J. Phys. Chem. B* 2009, **113**, 12154-12159.

106. T.K. Tam, G. Strack, M. Pita and E. Katz, *J. Amer. Chem. Soc.* **2009**, *131*, 11670-11671.

107. J. Zhou, G. Melman, M. Pita, M. Ornatska, X. Wang, A. Melman and E. Katz, *ChemBioChem* 2009, **10**, 1084-1090.

108. D. A. LaVan, T. McGuire and R. Langer, *Nature Biotechnology*, 2003, **21**, 1184-1191.

**Table 1.** Gate-function "quality measures" for the three-**AND**-gate network, derived from the data fits for the initially selected parameter set and after optimization of the network functioning. Results for two definitions of the output signal: at fixed gate-time and via the steady-state time-dependence slope, are shown; see Ref. 68 for details.

| Input Varied | Quality Measures[a,b] (the best possible value of these quantities is $\sqrt[4]{2} \approx 1.19$ ) | Slope Signal | Gate-Time Signal | Slope Signal | Gate-Time Signal | Gates Involved |
|---|---|---|---|---|---|---|
| | | *Initial experiment* | | *Optimized experiment* | | |
| $x_1$ | $\max[\sqrt{2}A_1 \ , \ 1/A_1]$ | 1.52[a] | 1.92[a] | 1.23 | 1.41 | 1 |
| $x_2$ | $\max[\sqrt{2}\sqrt{A_2 B_1} \ , \ 1/\sqrt{A_2 B_1}]$ | 2.04[a] | 1.56[a] | 1.41 | 1.41 | 1, 2 |
| $x_3$ | $\max[\sqrt{2}\sqrt[3]{A_3 B_2 B_1} \ , \ 1/\sqrt[3]{A_3 B_2 B_1}]$ | 1.26[a] | 1.21[a] | 1.21[b] | 1.24 | 1, 2, 3 |

[a] For the initial experiment, all the max[…] values were realized as inverses, rather than as products with $\sqrt{2}$ .

[b] For the optimized experiment, this single value was realized as an inverse; all the other values were $\sqrt{2}$….

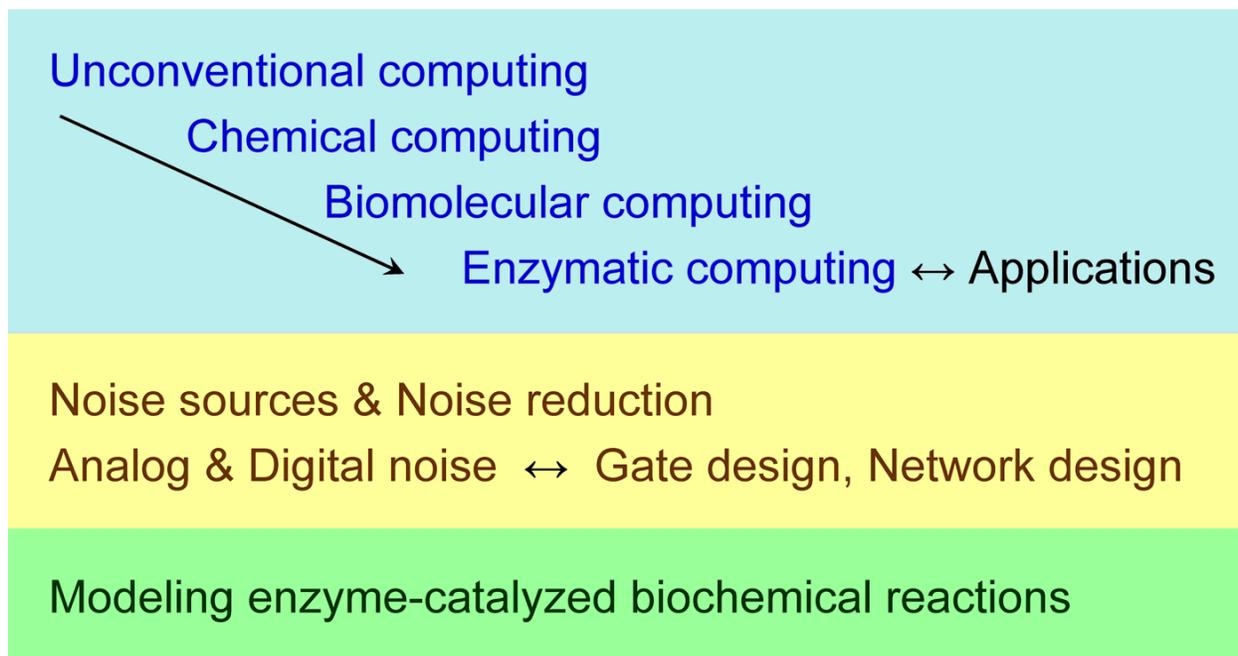

**Figure 1.** *Top:* Schematic of the interrelation of unconventional computing approaches, as addressed in this review. *Middle:* Topics in controlling the level of noise, which is a particularly important aspect for applications of enzyme-based information processing. *Bottom:* Approaches to modeling the relevant enzymatic reactions must be adjusted for proper analysis and design of "biocomputing" systems.

| Various paradigms for scalable information processing: | Examples of presently known systems/realizations: |
| --- | --- |
| ● Digital electronic circuitry | Modern electronic computers |
| ● Life | Living organisms |
| ● Quantum parallelism | Few-qubit register |
| ● "Ensemble" parallelism | DNA computing |

**Figure 2.** Schematic of the presently known paradigms for scalable, supposedly "fault-tolerant" information processing.

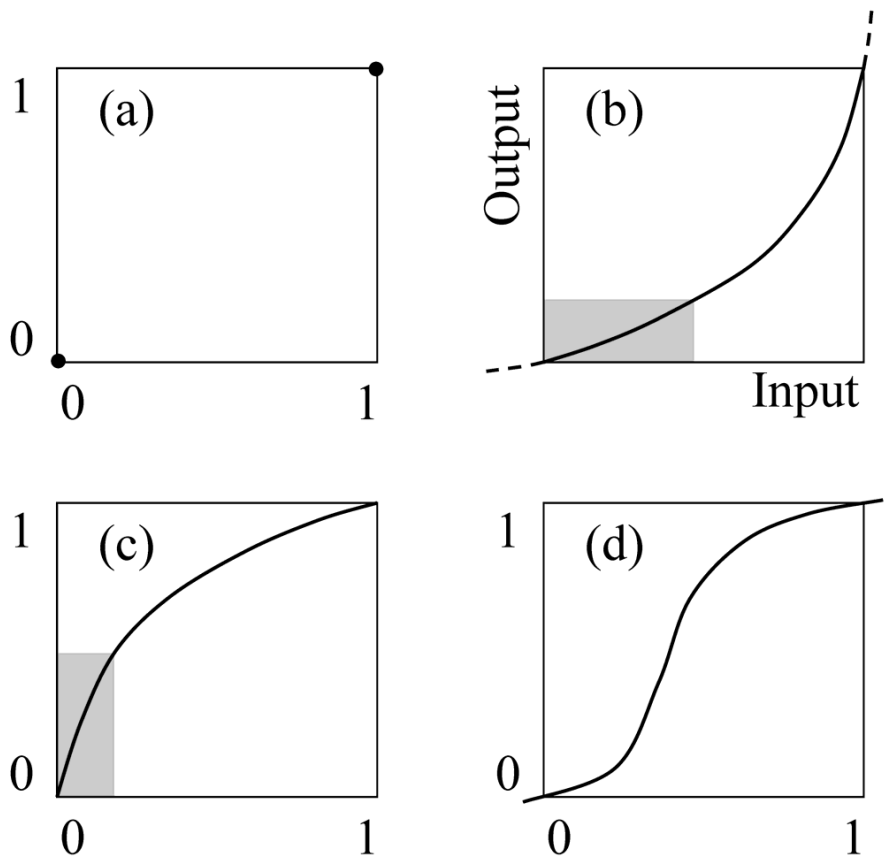

**Figure 3.** (a) The identity function mapping digital **0** and **1** to the same values, shown by the two bold dots in the figure. (b) A concave response curve with the shaded region indicating the approximately linear section utilized in "proportional" sensor applications. The input and output signals are not limited to the range bounded by the "digital" values and they do not have to be at the physical zero and some specific positive value. They can also be considered for values beyond the selected "digital" range: the broken-line sections. (c) The convex response typical of catalytic biochemical reactions. Here there is still an identifiable linear regime, typically near the physical zero concentrations, as well as the saturation regime for larger concentrations. The signals are shown normalized to the "digital" range between 0 and 1. Most studies of the (bio)chemical gates thus far have focused on the digital values, and more recently on the range between 0 and 1, ignoring the values beyond that range. (d) The sigmoid, filter-like response, which can (and should) also be considered above the digital value 1, as well as below the digital value 0 if it is not defined at the physical zero of concentrations. This is shown by the curve protruding beyond the "logic" interval.

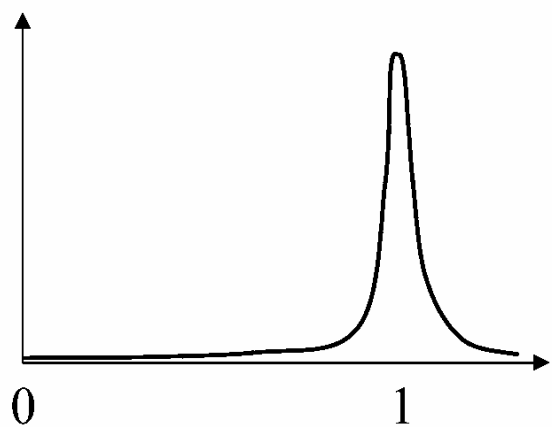

**Figure 4.** Spread of the signal about the expected digital-**1** result.

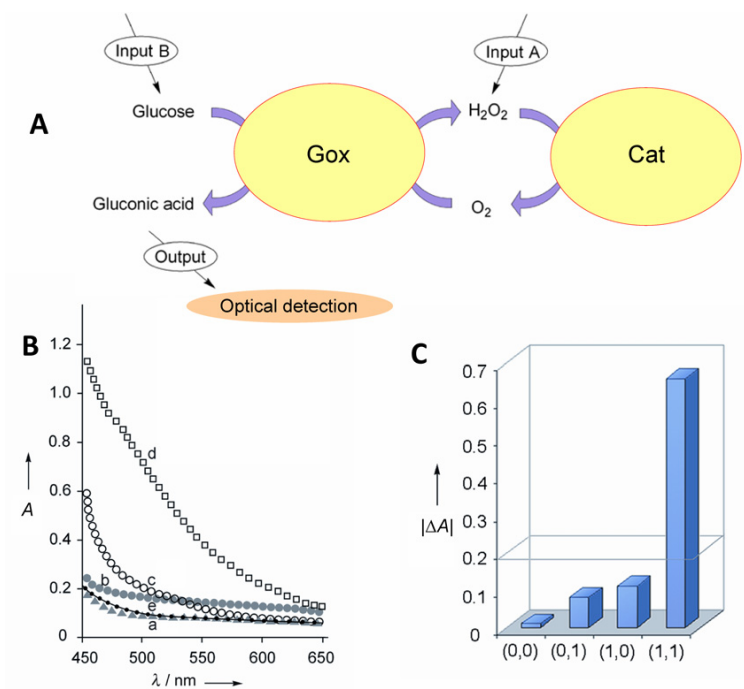

**Figure 5.** (A) Enzyme-based Boolean **AND** logic gate. (B) Optical changes in the system upon indirect analysis of gluconic acid[58] generated in the presence of different combinations of the input signals: (a) **0,0**; (b) **0,1**; (c) **1,0**; (d) **1,1**; and (e) prior to the signals application. (C) Bar diagram showing the absorbance changes,[58] $|\Delta A|$, derived from the spectra for various combinations of the input signals, and the threshold value (here 0.2) above which the digital **1** is registered.

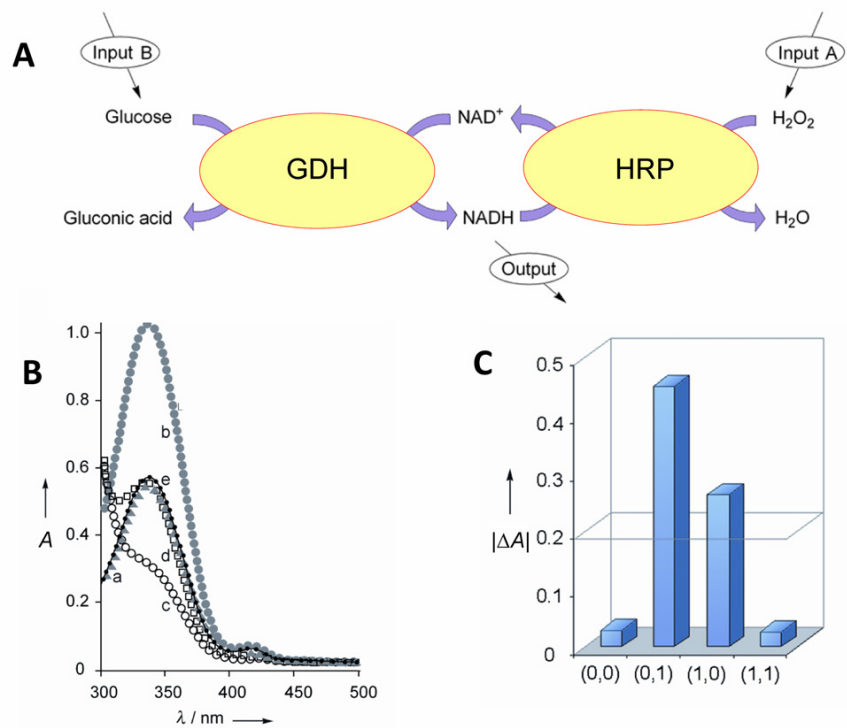

**Figure 6.** (A) Enzyme-based Boolean **XOR** logic gate. (B) Optical changes in the system generated[58] in the presence of different combinations of the input signals: (a) **0,0**; (b) **0,1**; (c) **1,0**; (d) **1,1**; and (e) prior to the signals application. (C) Bar diagram showing the absorbance changes,[58] $|\Delta A|$, for various combinations of the input signals, and the threshold value (0.2) above which the digital **1** is registered.

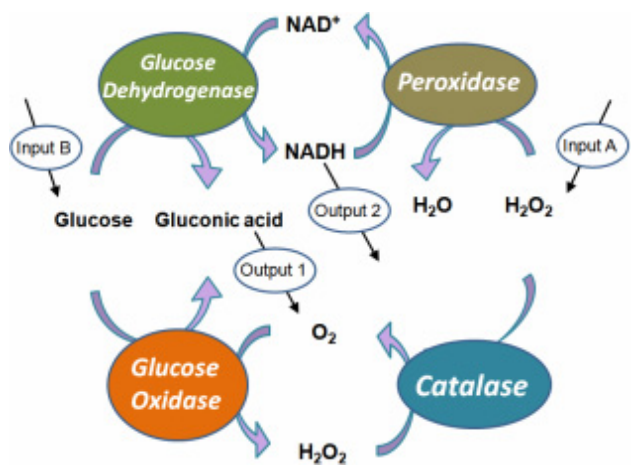

**Figure 7.** Enzyme based half-adder.[59]

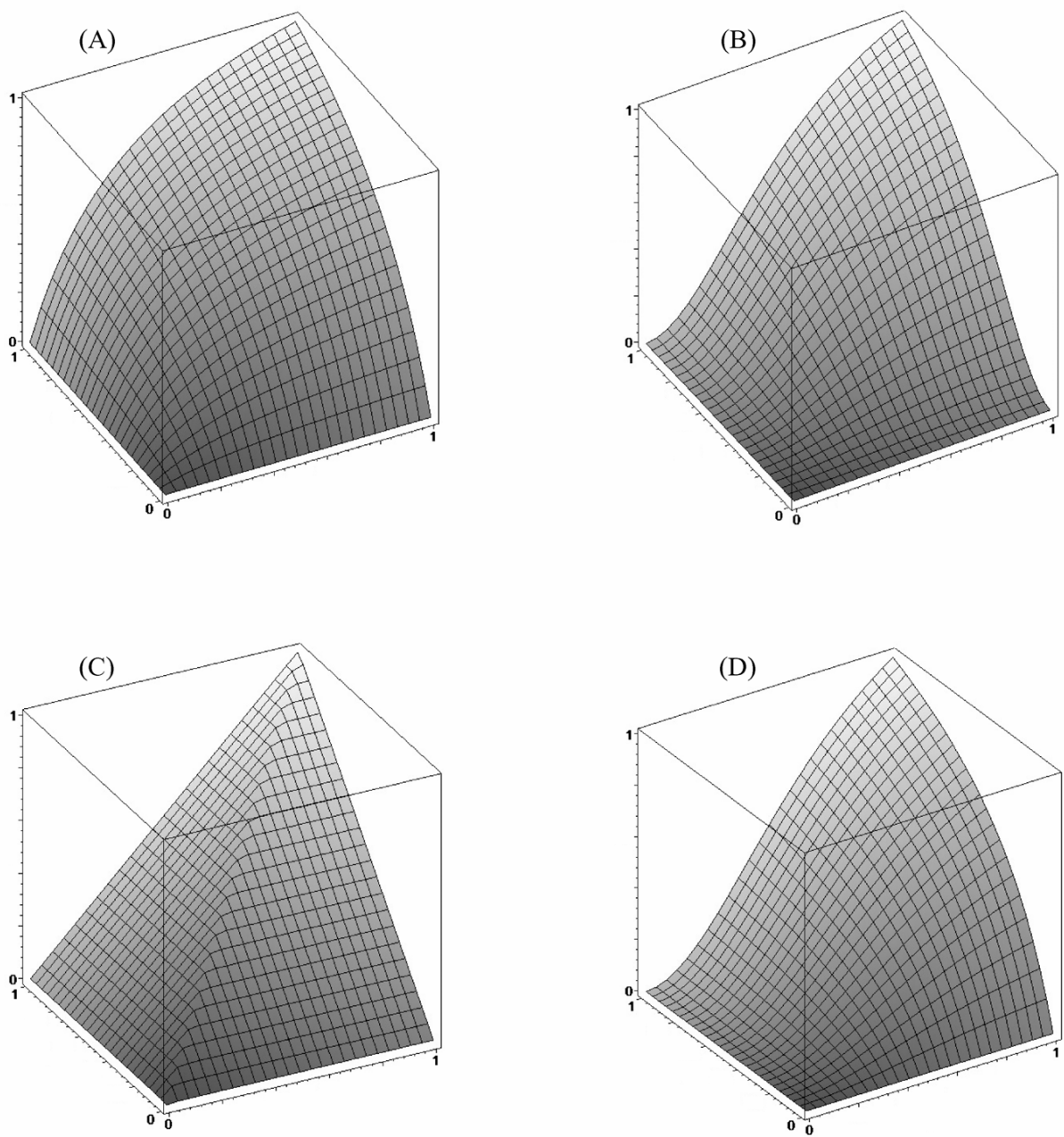

**Figure 8.** Two-input (one-output) analog-function surfaces: (A) Typical response surface for biocatalytic reactions. (B) Desirable response, sigmoid in both inputs. (C) Response surface that allows[73] elimination of noise amplification without sigmoid behavior. (D) Response surface sigmoid in one of the inputs,[2] which also avoids noise amplification for a proper choice of the parameters.

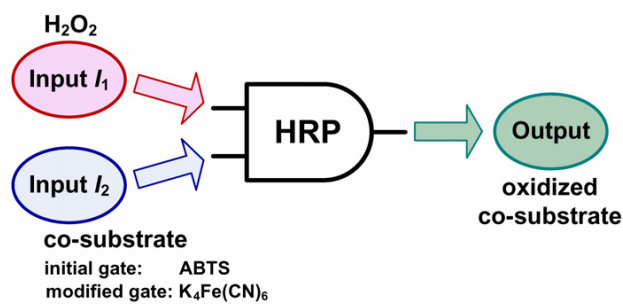

**Figure 9.** Schematics of the illustrative single-enzyme HRP-based **AND** logic gate.

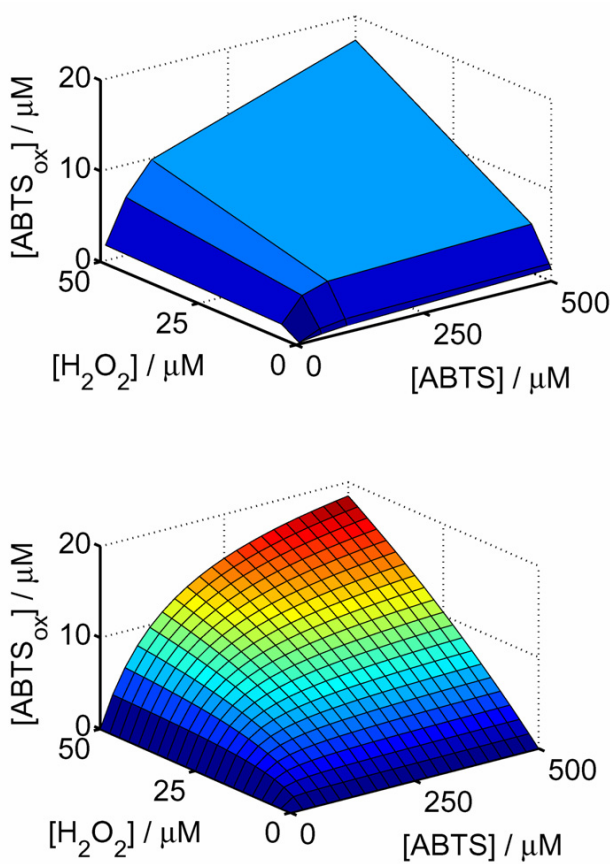

**Figure 10.** *Upper panel:* Experimental data for the **AND** gate carried out by HRP as the biocatalyst, with $H_2O_2$ and ABTS as the two inputs. The data are not scaled to the "logic" ranges in order to show the reference logic-**1** values. Details of the experiment, as well as values of additional parameters, can be found in Ref. 73. *Lower panel:* Theoretical data fit according to Eq. (3–4).

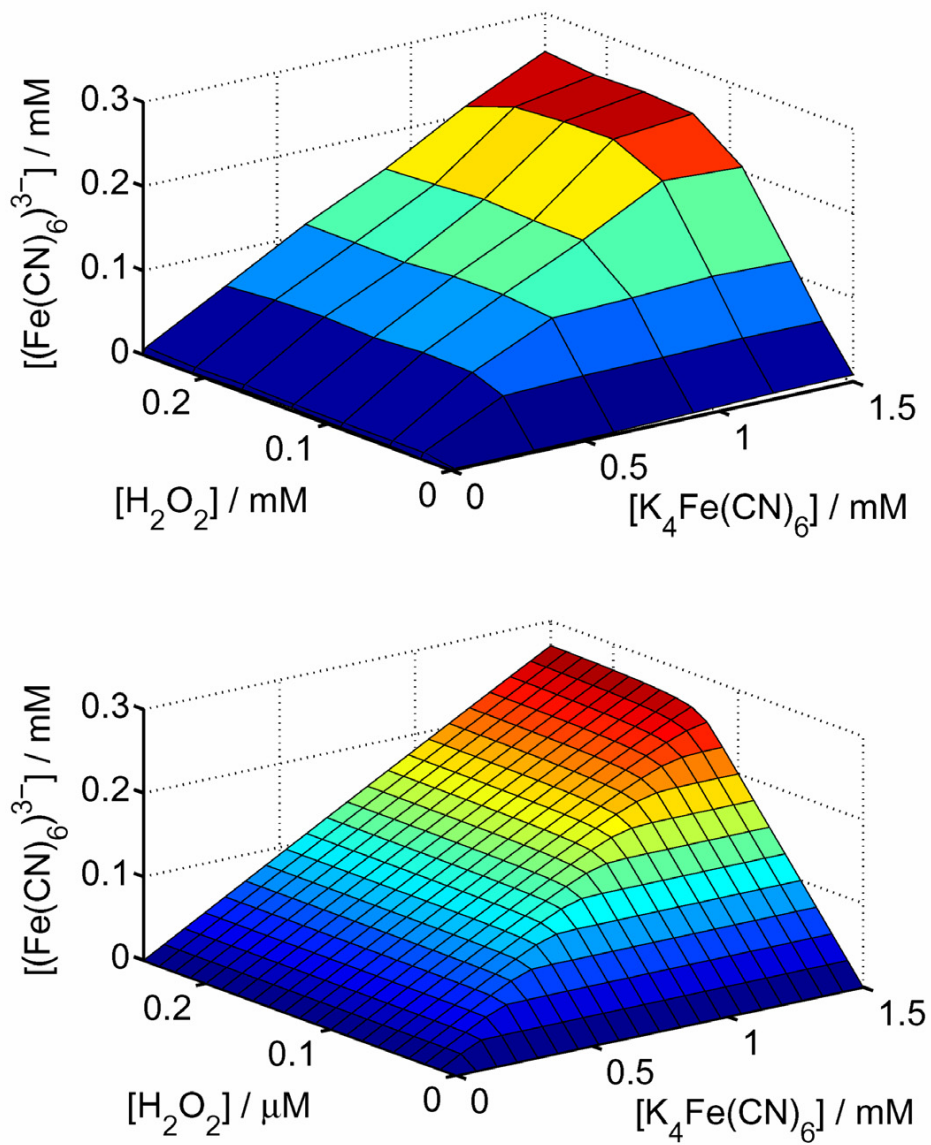

**Figure 11.** *Upper panel:* Experimental data for the **AND** gate carried out by HRP as the biocatalyst, with inputs H$_2$O$_2$ and ferrocyanide (introduced into solution as potassium ferrocyanide). The data are not scaled to the "logic" ranges in order to show the reference logic-**1** values (selected as experimentally convenient values, not related to those in Figure 10). Details of the experiment, as well as values of additional parameters, can be found in Ref. 73. *Lower panel:* The theoretical data fit according to Eq. (3–4).

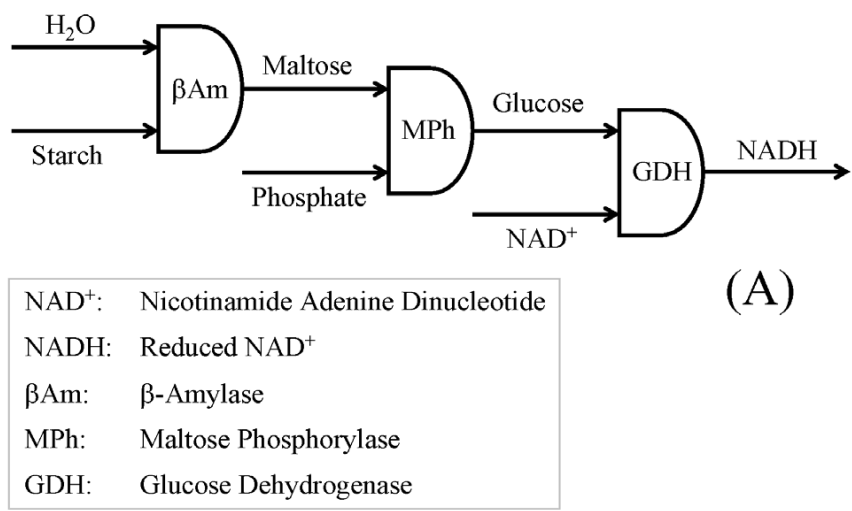

| NAD$^+$: | Nicotinamide Adenine Dinucleotide |
| NADH: | Reduced NAD$^+$ |
| βAm: | β-Amylase |
| MPh: | Maltose Phosphorylase |
| GDH: | Glucose Dehydrogenase |

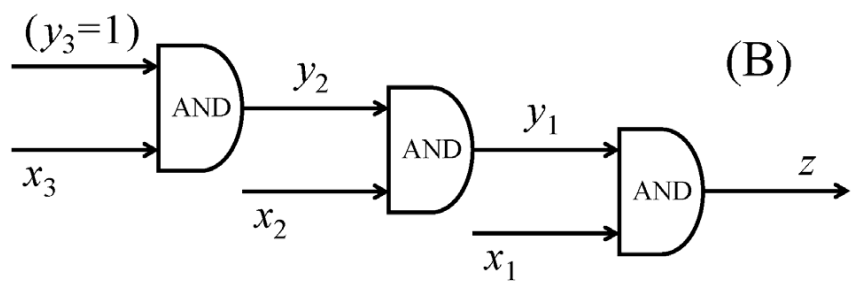

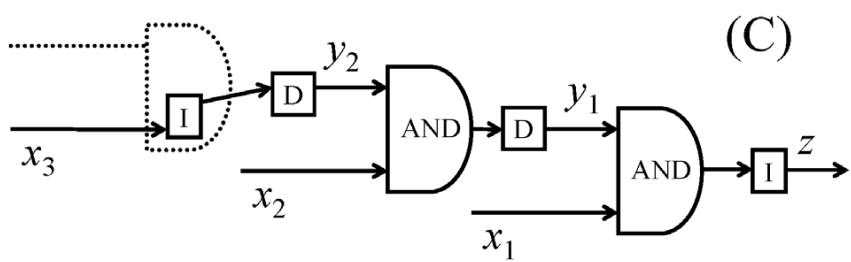

**Figure 12.** (A) Sequence of three enzyme-based biocatalytic reactions carried out in solution, with the output detected optically, as described in Ref. 68, where details of the experiment can be found. The inset offers a legend for the biochemical name abbreviations. (B) Modular representation of the realized sequence of enzymatic reactions as a network of three **AND** gates. (C) A more realistic, but not modular, network representation involving **AND**, **I**dentity and **D**elayed identity gates, as detailed in the text.

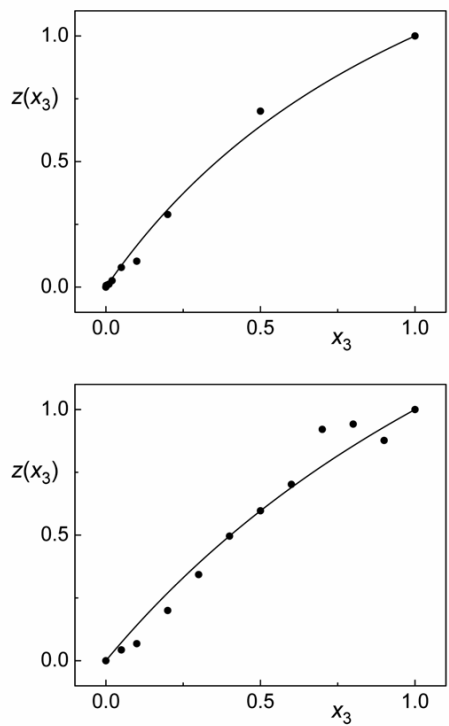

**Figure 13.** *Upper panel:* Experimental data for the optical-signal output at fixed time, $t^{\text{gate}}$, as a function of the input 3 (starch: cf. Figure 12), before any network optimization. The solid line is the data fit according to Eq. (8). *Lower panel:* The same after network optimization. Details of the experiments, as well as the values of the various parameters, can be found in Ref. 68. Note that all the variables here were rescaled to the "logic" ranges [0,1].

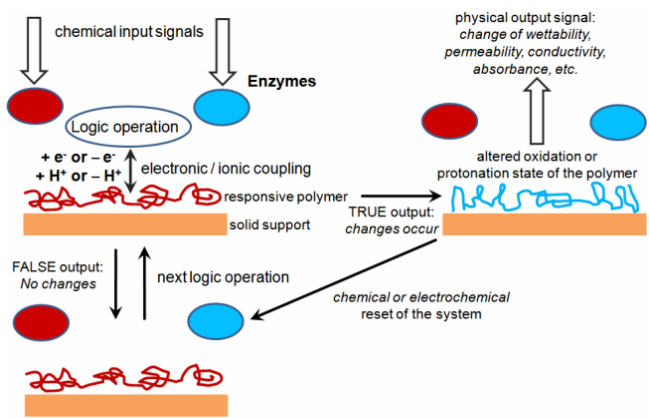

**Figure 14.** Scheme showing the general concept of interfacing enzyme-based logic systems with signal-responsive polymers operating as "smart" chemical actuators controlled by the gate output signals.

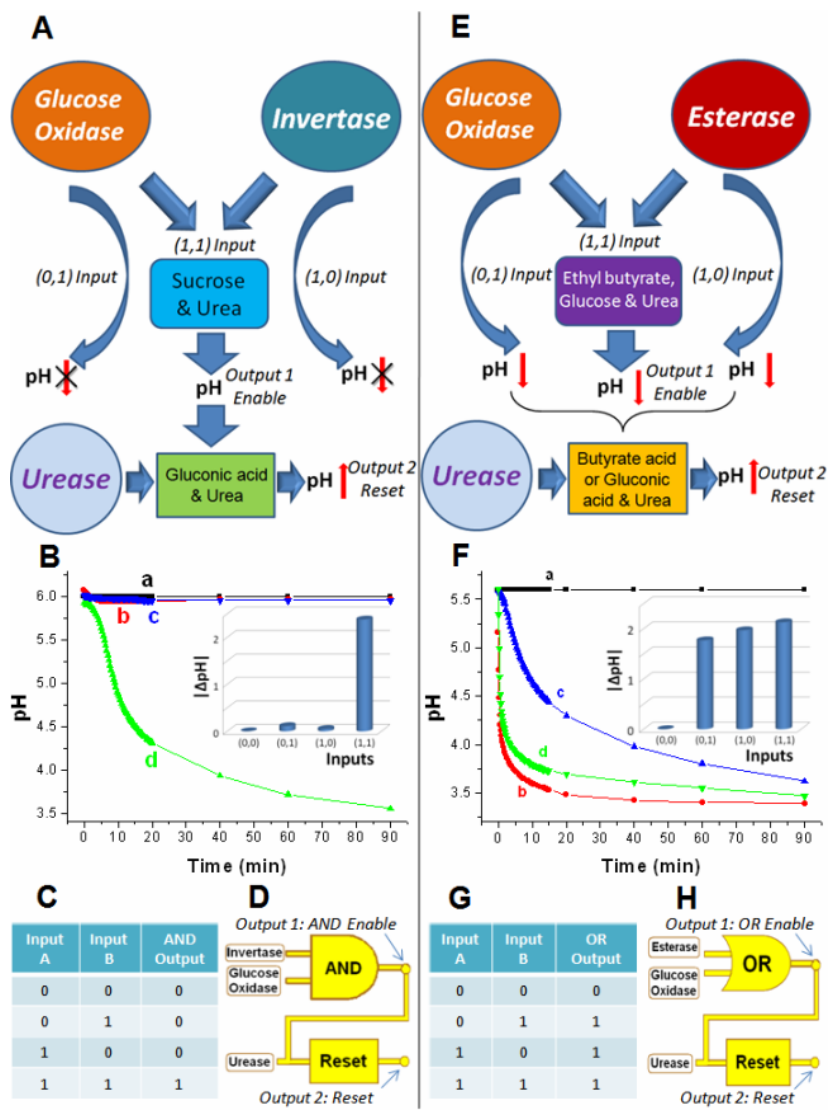

**Figure 15.** Biochemical logic gates with the enzymes used as input signals to activate the gate operation: the absence of the enzyme is considered as "**0**" and the presence as "**1**" input signals. The **Reset** function was catalyzed by urease. (A) The **AND** gate based on GOx and Inv catalyzed reactions. (B) pH-changes generated *in situ* by the **AND** gate upon different combinations of the input signals: a) "**0,0**", b) "**0,1**", c) "**1,0**" and d) "**1,1**". Inset: Bar diagram showing the pH changes as the output signals of the **AND** gate. (C) The truth table of the **AND** gate showing the digital output signals corresponding to the pH changes generated upon different combinations of the input signals. (D) Equivalent electronic circuit for the biochemical **AND-Reset** logic operations. (E) The **OR** gate based on GOx and Est catalyzed reactions. (F-H) The same as (B-D) for the **OR** gate.

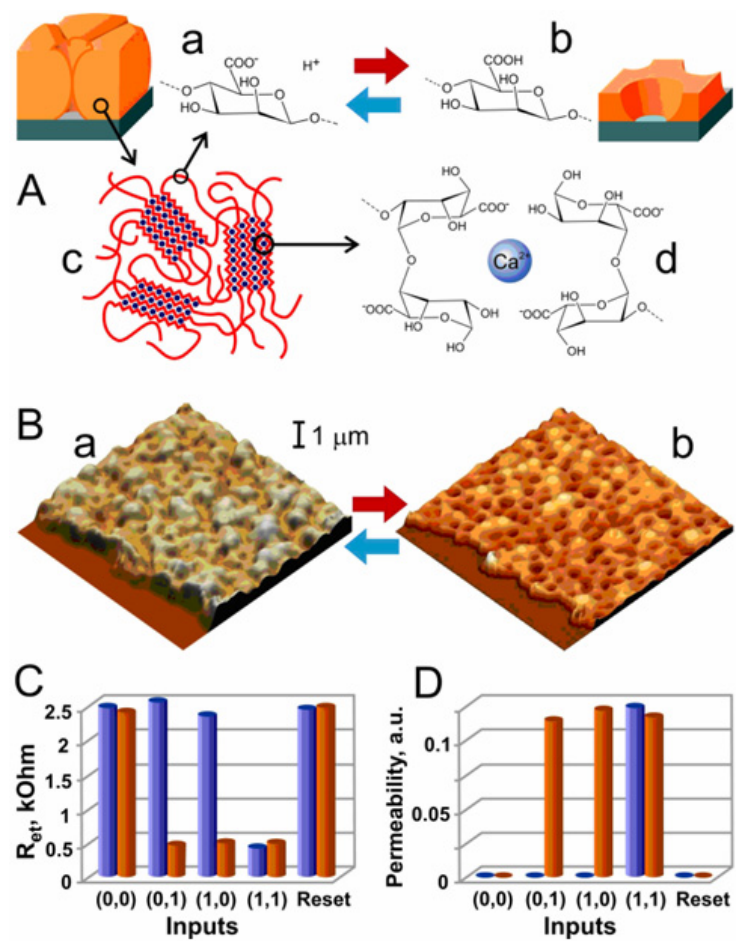

**Figure 16.** Signal-responsive membrane coupled with enzyme-based logic gates. (A) Schematic representations of a single pore of the polyelectrolyte membrane switched between the closed (a) and open (b) states. (c) The structure of the alginate hydrogel constituted of D-mannuronic acid and L-guluronic acid residues cross-linked with divalent ions ($Ca^{2+}$) in (d) an egg-box-like conformation. The swelling and shrinking of the hydrogel is attributed to the ionization (a) and protonation (b) of the unbound carboxyl groups at pH > 5 and pH < 4, respectively. (B) SPM topography images (10×10 $\mu m^2$) of the swollen (a) and shrunken (b) membrane. (C) Electron transfer resistance, $R_{et}$, of the membrane deposited on the electrode surface derived from the impedance spectroscopy measurements obtained upon different combinations of input signals. (D) Relative permeability (ratio of the membrane permeability deposited on the supporting filter to the permeability of the filter with no membrane) for Rhodamine B obtained upon different combinations of the input signals. Blue and red bars correspond to the **AND** / **OR** gates, respectively (a.u. denotes arbitrary units).

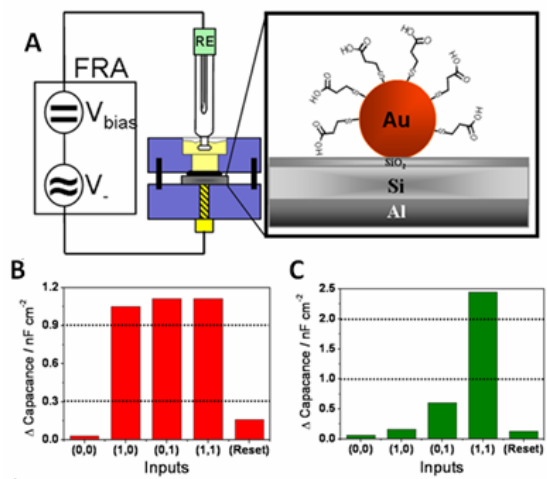

**Figure 17.** Electronic scheme (A) of the signal-transducing device based on a Si-chip modified with Au nanoparticles coated with a pH-sensitive organic shell (FRA = frequency response analyzer; RE = reference electrode). The bar diagrams showing the output signals generated by **OR** (B) / **AND** (C) enzyme logic gates and transduced by the Si-chip in the form of capacitance changes. The dashed lines correspond to the threshold values: the output signals below the first threshold were considered as **0**, while the signals above the second threshold were treated as **1**. (Adapted from Ref. 96 with permission.

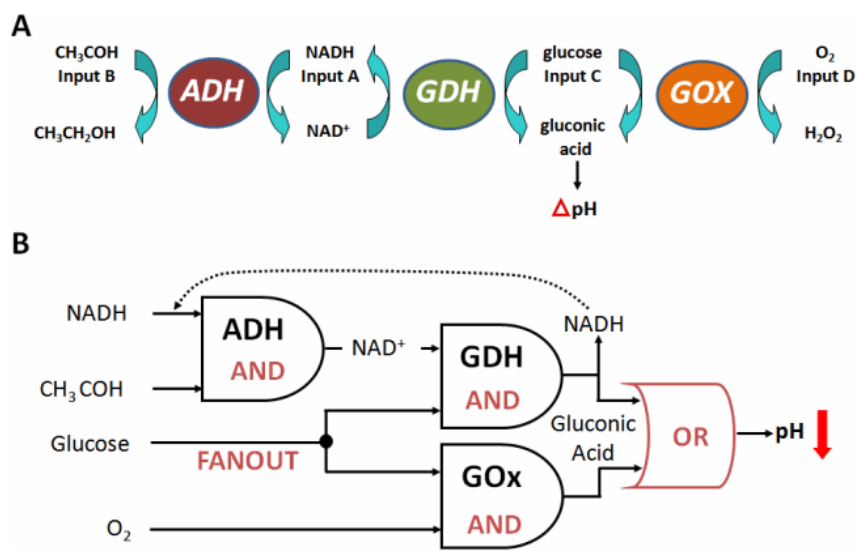

**Figure 18.** (A) Multi-gate / multi-signal processing enzyme logic system producing *in situ* pH changes as the output signal. (B) The equivalent logic circuitry for the biocatalytic cascade.

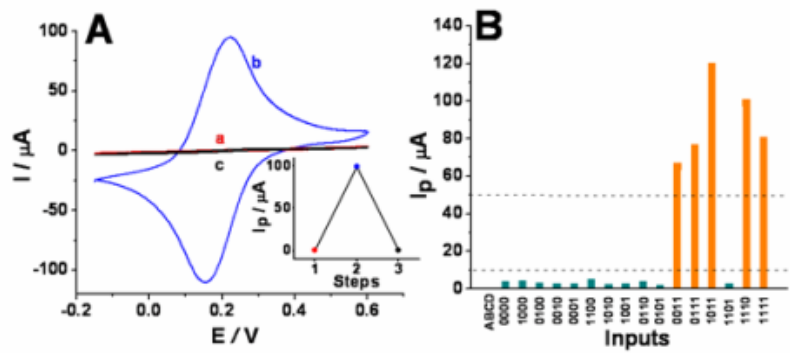

**Figure 19.** Electrochemical transduction of the chemical signals processed by the enzyme logic network shown in Figure 18. (A) Cyclic voltammograms obtained for the ITO electrode modified with the P4VP-polymer brush in: a) the initial OFF state, pH ca. 6.7, b) ON state enabled by the input combinations resulting in acidifying the solution to pH ca. 4.3, and c) *in situ* reset to the OFF state, pH ca. 8.8. Inset: reversible current changes upon switching the electrode ON-OFF. The dotted lines show threshold values separating logic **1**, undefined, and logic **0** output signals. (See Ref. 98 for the details of the experiment.)

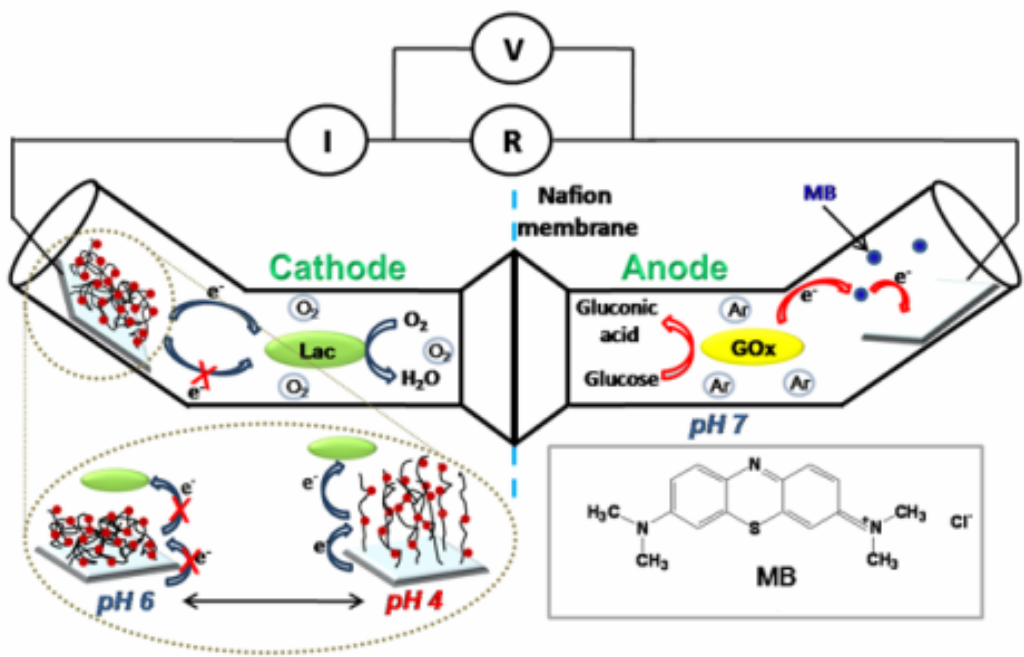

**Figure 20.** Biofuel cell composed of the pH-switchable logically controlled biocatalytic cathode and glucose oxidizing anode.

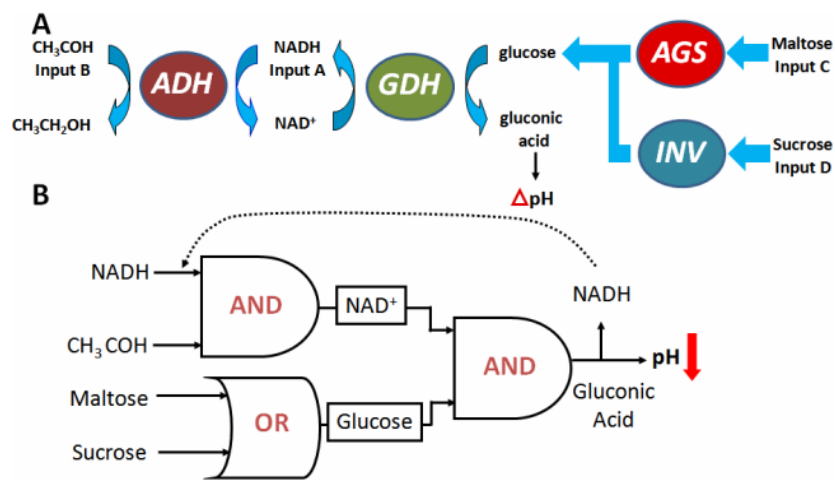

**Figure 21.** (A) The cascade of reactions biocatalyzed by alcohol dehydrogenase (ADH), amyloglucosidase (AGS), invertase (INV) and glucose dehydrogenase (GDH) and triggered by chemical input signals: NADH, acetaldehyde, maltose and sucrose, added in different combinations. (B) The logic network composed of three concatenated gates, corresponding to the cascade of enzymatic reactions in (A).

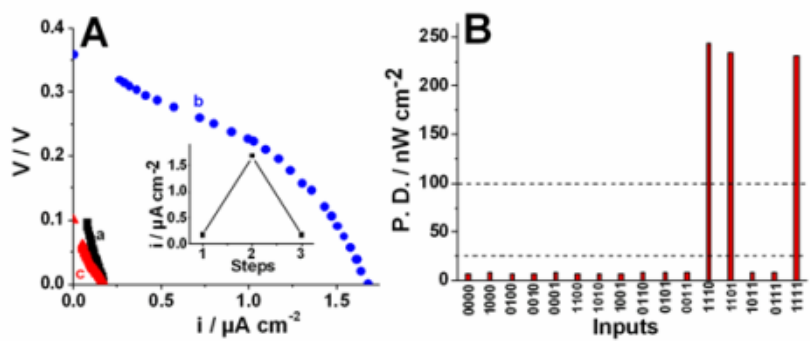

**Figure 22.** (A) V-i polarization curves obtained for the biofuel cell with different load resistances: (a) in the inactive state prior to the addition of the biochemical input signals (pH value in the cathodic compartment ca. 6); (b) in the active state after the cathode was activated by changing pH to ca. 4.3 by the biochemical signals; (c) after the **Reset** function activated by the addition of 5 mM urea. Inset: Switchable $i_{sc}$ upon transition of the biofuel cell from the mute to active state and back, performed as biochemical signals are being processed by the enzyme logic network. (B) The bar diagram showing the power density produced by the biofuel cell in response to different patterns of the chemical input signals. Dashed lines show thresholds separating digital **0**, undefined, and **1** output signals produced by the system.